\documentclass[aps,pra,twocolumn,amssymb,amsfonts,floatfix,showpacs]{revtex4}

\usepackage{graphicx}
\usepackage{dcolumn}
\usepackage{bm}
\usepackage{mathbbol}
\begin{document}

\title{Dynamical control of qubit coherence:
Random versus deterministic schemes}

\author{Lea F. Santos}
\email{Lea.F.Dos.Santos@Dartmouth.edu}
\author{Lorenza Viola}
\email{Lorenza.Viola@Dartmouth.edu}
\affiliation{\mbox{Department of Physics and Astronomy, 
Dartmouth College, 6127 Wilder Laboratory, Hanover, NH 03755, USA}}

\date{\today}

\begin{abstract}
We revisit the problem of switching off unwanted phase evolution and
decoherence in a single two-state quantum system in the light of
recent results on random dynamical decoupling methods [L. Viola and
E. Knill, Phys.  Rev. Lett. {\bf 94}, 060502 (2005)].  A systematic
comparison with standard cyclic decoupling is effected for a variety
of dynamical regimes, including the case of both semiclassical and
fully quantum decoherence models. In particular, exact analytical
expressions are derived for randomized control of decoherence from a
bosonic environment.  We investigate quantitatively control protocols
based on purely deterministic, purely random, as well as hybrid
design, and identify their relative merits and weaknesses at improving
system performance.  We find that for time-independent systems, hybrid
protocols tend to perform better than pure random and may improve over
standard asymmetric schemes, whereas random protocols can be
considerably more stable against fluctuations in the system
parameters.  Beside shedding light on the physical requirements
underlying randomized control, our analysis further demonstrates the
potential for explicit control settings where the latter may
significantly improve over conventional schemes.
\end{abstract}

\pacs{03.67.Pp, 03.65.Yz, 05.40.Ca, 89.70.+c}

\maketitle

\section{Introduction}

The design and characterization of strategies for controlling quantum
dynamics is vital to a broad spectrum of applications within contemporary
physics and engineering. These range from traditional coherent-control
settings like high-resolution nuclear~\cite{HaeberlenBook,ErnstBook}
and molecular spectroscopy~\cite{BrumerBook}, to a variety of tasks
motivated by the rapidly growing field of quantum information
science~\cite{NielsenBook}.  In particular, the ability to counteract
decoherence effects that unavoidably arise in the dynamics of a
real-world quantum system coupled to its surrounding environment is 
a prerequisite for scalable realizations of quantum information
processing (QIP), as actively pursued through a variety of proposed
device technologies~\cite{QCRoadmap}.

Active decoupling techniques offer a conceptually simple yet powerful
control-theoretic setting for quantum-dynamical engineering of both
closed-system (unitary) and open-system (non-unitary) evolutions.
Inspired by the idea of {\em coherent averaging} of interactions by
means of tailored pulse sequences in nuclear magnetic resonance (NMR)
spectroscopy~\cite{Waugh}, decoupling protocols consist of repetitive
sequences of control operations (typically drawn from a finite
repertoire), whose net effect is to coherently modify the natural
target dynamics to a desired one.  In practice, a critical decoupling
task is the selective removal of unwanted couplings between subsystems
of a fully or partially controllable composite quantum system.
Historically, a prototype example is the elimination of unwanted phase
evolution in interacting spin systems via trains of $\pi$-pulses (the
so-called Hahn-echo and Carr-Purcell
sequences~\cite{Hahn-Echo,CP-Echo}).  For open quantum systems, this
line of reasoning motivates the question of whether removing the
coupling between the system of interest and its environment may be
feasible by a control action restricted to the former only.  Such a
question was addressed in~\cite{Viola1998} for the paradigmatic case
of a single qubit coupled to a bosonic reservoir, establishing the
possibility of decoherence suppression in the limit of rapid
spin flipping via the echo sequence mentioned above.

The study of dynamical decoupling as a general strategy for quantum
coherent and error control has since then attracted a growing interest
from the point of view of both model-independent decoupling design and
optimization, and the application to specific physical systems.
Representative contributions include the extension to arbitrary
finite-dimensional systems via
dynamical-algebraic~\cite{Viola1,ZanardiSym0},
geometric~\cite{Byrd2002b}, and linear-algebraic~\cite{Ticozzi05}
formulations; the construction of fault-tolerant
Eulerian~\cite{Viola2003Euler} and concatenated decoupling
protocols~\cite{Khodjasteh2004}, as well as efficient combinatorial
schemes~\cite{Jones,Stoll,Leung02,Wocjan}; the connection with quantum
Zeno physics~\cite{Facchi2004}; proposed applications to the
compensation of specific decoherence mechanisms (notably, magnetic
state decoherence~\cite{Search} and 1/$f$
noise~\cite{Shiokawa2002,Gutmann2003,Falci2004,Faoro2004,Gutmann2004})
and/or the removal of unwanted evolution within
trapped-ion~\cite{Tombesi,Lidar-Ion} and solid-state quantum computing
architectures~\cite{ss}.  These theoretical advances have been
paralleled by steady experimental progress. Beginning with a
proof-of-principle demonstration of decoherence suppression in a
single-photon polarization interferometer~\cite{Berglund2000},
dynamical decoupling techniques have been implemented alone and in
conjunction with quantum error correction within liquid-state NMR
QIP~\cite{Cory-Overview,BoulantDec}, and have inspired
charge-based~\cite{Nakamura2002} and flux-based~\cite{Chiorescu} echo
experiments in superconducting qubits.  Recently, dynamic decoherence
control of a solid-state nuclear quadrupole qubit has been
reported~\cite{Fraval}.

All the formulations of dynamical decoupling mentioned so far share
the feature of involving purely {\em deterministic} control actions.
In the simplest setting, these are arbitrarily strong, effectively
instantaneous rotations (so-called {\em bang-bang controls}) chosen
from a discrete group ${\cal G}$.  Decoupling according to ${\cal G}$
is then accomplished by sequentially cycling the control propagator
through {\em all} the elements of ${\cal G}$.  If $\Delta t$ denotes
the separation between consecutive control operations, this translates
into a minimal averaging time scale $T_c=|{\cal G}|\Delta t$, of
length proportional to the size $|\cal G|$ of $\cal G$.

The exploration of decoupling schemes incorporating {\em stochastic}
control actions was only recently undertaken. A general
control-theoretic framework was introduced by Viola and Knill
in~\cite{Viola2005Random} (see also~\cite{Violacdc05}), based on the
idea of seeking faster convergence (with respect to an appropriately
defined metric) by {\em randomly sampling} rather than systematically
implementing control operations from ${\cal G}$.  Based on general
lower bounds for pure-state error probabilities, the analysis
of~\cite{Viola2005Random} indicated that random schemes could
outperform their cyclic counterpart in situations where a large number
of elementary control operations is required or, even for small
control groups, when the interactions to be removed vary themselves in
time over time scales long compared to $\Delta t$ but short compared
to $T_c$.  Furthermore, it also suggested that advantageous features
of pure cyclic and random methods could be enhanced by appropriately
merging protocols within a hybrid design.  The usefulness of
randomization in the context of actively suppressing {\em coherent}
errors due to residual static interactions was meanwhile independently
demonstrated by the so-called {\em Pauli Random Error Correction
Method} (PAREC), followed by the more recent {\em Embedded Dynamical
Decoupling Method} -- both due to Kern and
coworkers~\cite{Kern2004,Kern2005}.  Both protocols may be
conceptually understood as following from randomization over the Pauli
group ${\cal G}_P=\{ \openone, \sigma_x, \sigma_y, \sigma_z\}$, used
alone or, respectively, in conjunction with a second set of
deterministic control operations.

Our goal in this work is twofold: first, to develop a quantitative
understanding of {\em typical} randomized control performance for both
{\em coherent and decoherent phase errors}, beginning from the
simplest scenario of a single qubit already investigated in detail in
the deterministic case~\cite{Viola1998}; second, to clarify the {\em
physical} picture underlying random control, by devoting, in
particular, special attention to elucidate the control action and
requirements in rotating frames associated with different dynamical
representations.  The fact that the controlled dynamics remains
exactly solvable in the bang-bang (BB) limit makes the single-qubit
pure-dephasing setting an ideal test-bed for these purposes.  From a
general standpoint, since spin-flip decoupling corresponds to
averaging over the smallest (nontrivial) group ${\cal Z}_2=\{0,1\}$,
with $T_c=2\Delta t$~\cite{Viola1,ZanardiSym0}, this system is not yet
expected to show the full advantage of the random approach.
Remarkably, however, control scenarios can still be identified, where
randomized protocols indeed represent the most suitable choice.

The content of the paper is organized as follows.  After laying out
the relevant system and control settings in Sect.~II, we begin the
comparison between cyclic and randomized protocols by studying the
task of phase refocusing in a qubit evolving unitarily in Sect.~III.
Control of decoherence from purely dephasing semiclassical and quantum
environments is investigated in the main part of the paper, Sects.~IV
and V.  We focus on the relevant situations of decoherence due to
random telegraph noise and to a fully quantum bosonic bath,
respectively.  Both exact analytical and numerical results for the
controlled decoherence process are presented in the latter case.  We
summarize our results and discuss their significance from the broader
perspective of constructively exploiting randomness in physical
systems in Sect.~VI, by also pointing to some directions for future
research.  Additional technical considerations are included in a
separate Appendix.

\section{Single-qubit quantum-control settings}

Our target system $S$ is a single qubit, living on a state space
${\cal H}_S\simeq {\mathbb C}^2$. The influence of the surrounding
environment may be formally accounted for by two main modifications to
the isolated qubit dynamics. First, $S$ may couple to effectively {\em
classical} degrees of freedom, whose net effect may be modeled through
a deterministic or random time-dependent modification of the system
parameters.  Additionally, $S$ may couple to a {\em quantum}
environment $E$, that is, a second quantum system defined on a state
space ${\cal H}_E$ with which $S$ may become entangled in the course
of the evolution.  For the present purposes, $E$ will be schematized
as a bosonic reservoir consisting of independent harmonic modes.  Let
$\openone_{S,E}$ denote the identity operator on ${\cal H}_{S,E}$,
respectively. Throughout the paper, we will consider different
dynamical scenarios, corresponding to special cases of the following
total drift Hamiltonian on ${\cal H}_S \otimes {\cal H}_E$:
\begin{eqnarray}
&&H_0(t)=H_S(t)\otimes\openone_E + \openone_S\otimes H_E + H_{SE}(t)\:, 
\label{drift0}
\end{eqnarray}
where
\begin{equation}
\label{drift}
\left\{
\begin{array}{l}
H_S(t)=\frac{\omega_0(t)}{2}\sigma_z \:,  \\
\\
H_E = \sum_{k} \omega_k b^{\dagger}_k b_k \:,  \\
\\
H_{SE}(t)=\mu \, \sigma_z\otimes
\sum_k \Big(g_k (t) b^{\dagger}_k + g^{*}_k (t)b_k\Big) \:. 
\end{array}
\right.
\end{equation}
Here, we set $\hbar=1$, and $\sigma_i$ ($i=x,y,z$), $b^{\dagger}_k$
and $b_k$ denote Pauli spin matrices, and canonical creation and
annihilation bosonic operators of the $k$th environmental mode with
frequency $\omega_k$, respectively.  $\omega_0(t)$ and $g_k(t)$ are
real and complex functions that account for an effectively
time-dependent frequency of the system and its coupling to the $k$th
reservoir mode, respectively.  We shall write 
\begin{eqnarray}
\omega_0(t)& =&  \omega_0 +\delta \omega_0(t)\:, \nonumber \\
g_k(t) &=& g_k +\delta g_k (t)\:, 
\end{eqnarray}
for an appropriate choice of central values $\omega_0$, $g_k$ and
modulation functions $\delta \omega_0$, $\delta g_k$, respectively.
The adimensional parameter $\mu$ is introduced for notational
convenience, allowing to include ($\mu=1$) or not ($\mu=0$) the
coupling to $E$ as desired.  Physically, because $H_S(t)$ and
$H_{SE}(t)$ commute at all times, the above Hamiltonian describes a
{\em purely decohering} coupling between $S$ and $E$, which does not
entail energy exchange.  While in general dissipation might also
occur, focusing on pure decoherence is typically justified for
sufficiently short time scales~\cite{Palma,Breuer} and, as we shall
see, has the advantage of making exact solutions available as
benchmarks.

Control is introduced by adjoining a classical controller acting on
$S$, that is by adding a time-dependent term to the above target
Hamiltonian,
\begin{eqnarray}
H_0(t) \mapsto H_0(t) + H_c(t) \otimes \openone_E\:.
\end{eqnarray}
In our case, $H_c(t)$ will be designed so as to implement appropriate
sequences of BB pulses.  This may be accomplished by starting from a
rotating radiofrequency field (or, upon invoking the rotating wave
approximation, by a linearly-polarized oscillating field), described
by the following amplitude- and phase-modulated Hamiltonian:
$$ H_c(t)= 
\sum_{j} V^{(j)} (t)
\Big(\hspace*{-1mm}\cos [\omega t + \varphi_j (t)] \sigma_x + 
\sin[\omega t + \varphi_j (t)] \sigma_y \hspace*{-.5mm}\Big) \,, $$ 
with
\[V^{(j)}(t)=V[\Theta (t-t_j) - \Theta (t-t_j-\tau)]\:. \]
Here, $\Theta(\cdot)$ denotes the Heaviside step function (defined as
$\Theta(x)=0$ for $x\leq 0$ and $\Theta=1$ for $x>0$), $V$ and $\tau$
are positive parameters, and $t_j$ denotes the instants at which the
pulses are applied.  If the carrier frequency is tuned on resonance
with the central frequency, $\omega = \omega_0$, and the phase
$\varphi_j(t) = -\omega_0 t_j$ for each $j$, the above Hamiltonian 
schematizes a train of {\em identical} control pulses of amplitude $V$
and duration $\tau$ in the physical frame.  Under the BB requirement
of impulsive switching ($\tau \rightarrow 0$) with unbounded strength
($V\rightarrow \infty $), it is legitimate to neglect $H_0(t)$
(including possible off-resonant effects) within each pulse,
effectively leading to qubit rotations about the $\hat{x}$-axis.  In
particular, a $\pi$ rotation corresponds to $2V \tau =\pm \pi$ (see
also Appendix A).

In what follows, we shall focus on using trains of BB $\pi$-pulses to
effectively achieve a net evolution characterized by the identity
operator (the so-called {\em no-op} gate).  This requires averaging
unwanted (coherent or decoherent) $\sigma_z$ evolution generated by
either $H_S(t)$ or $H_{SE}(t)$ or both, by subjecting the system to
repeated spin-flips.  In group-theoretic terms such protocols have, as
mentioned, a transparent interpretation as implementing an average
over the group ${\cal Z}_2$, represented on ${\cal H}_S$ as $\hat{\cal
G}=\{ \hat{g}_\ell\}=\{ \openone, \sigma_x \}$~\cite{Viola1}.  The
quantum operation effecting such group averaging is the projector
$\Pi_{\cal G}$ on the space of operators commuting with $\hat{\cal
G}$, leading to
$$ \Pi_{\cal G}(\sigma_z) ={1\over |\cal G|} \sum_{g_\ell\in \cal G}
\hat{g_\ell}^\dagger \sigma_z \hat{g_\ell} = {1 \over 2} \Big(
\openone \sigma_z \openone + \sigma_x \sigma_z \sigma_x\Big) =0\:.
$$ Essentially, in {\em cyclic} decoupling schemes based on ${\cal G}$
the above symmetrization is accomplished through a {\em time} average
of the effective Hamiltonian determining the evolution over a cycle,
$T_c$; in {\em random} schemes, it emerges from an {\em ensemble}
average over different control histories, taken with respect to the
uniform probability measure over ${\cal
G}$~\cite{Viola2005Random,haar}.  Neither deterministic nor stochastic
sequences of $\pi$ pulses achieve an exact implementation of $
\Pi_{\cal G}$ for a fully generic Hamiltonian as in
Eqs.~(\ref{drift0})-(\ref{drift}), except in the ideal limit of
arbitrarily fast control where the separation between pulses
approaches zero.  Therefore, it makes sense to compare the performance
attainable by different control sequences for realistic control
rates. In this paper we shall focus on the following options.

\begin{figure}[tb]
\includegraphics[width=3.in]{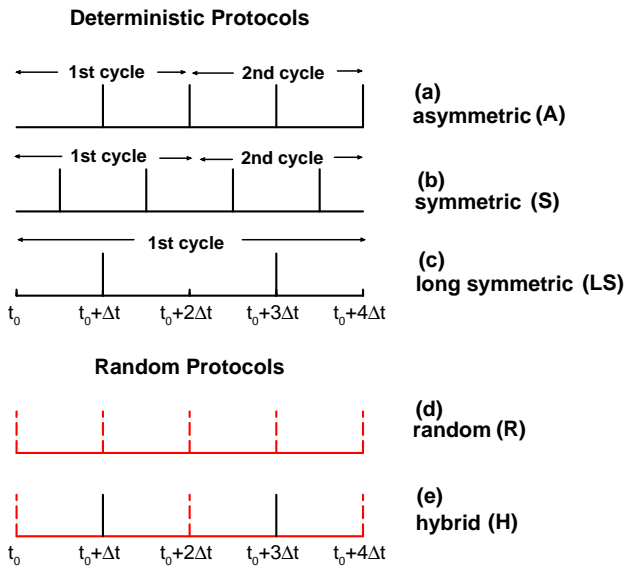}
\caption{Pictorial representation of relevant control protocols used
for coherence control.  Deterministic pulses are indicated with full
vertical lines, while random pulses correspond to dashed vertical
lines.}
\label{fig:scheme}
\end{figure}

$\bullet$ {\bf Asymmetric cyclic protocol (A)},
Fig.~\ref{fig:scheme}(a).  This is the protocol used
in~\cite{Viola1998}, corresponding to repeated spin-echoes.  Cyclicity
is ensured by subjecting the system to an even number of equally
spaced $\pi$-pulses, applied at $t_j=t_0 + j\Delta t$, $j =1,2,
\ldots$, in the limit $\tau \rightarrow 0$.  The elementary cycle
consists of two pulses: the first one, applied after the system
evolved freely for an interval $\Delta t$, reverses the qubit original
state and the second one, applied a time $\Delta t$ later, restores
its original state.

$\bullet$ {\bf Symmetric cyclic protocol (S)},
Fig.~\ref{fig:scheme}(b).  This protocols, which is directly inspired
to the Carr-Purcell sequence of NMR, is obtained from (A) by
rearranging the two $\pi$-pulses within each cycle in such a way that
the control propagator is symmetric with respect to the middle point.
The first pulse is applied at $t_1=t_0+\Delta t/2$ and the next ones
at $t_j=t_1 + (j-1)\Delta t$, with $j > 1$.  Both the A- and
S-protocols have a cycle time $T_c=2\Delta t$, and lead to the same
averaging in the limit $\Delta t \rightarrow 0$.  For finite $\Delta
t$, however, the symmetry of the S-protocol guarantees the
cancellation of lowest-order corrections $O(\Delta t)$, resulting in
superior averaging ~\cite{ErnstBook,Gheorghiu,Faoro2004}.

$\bullet$ {\bf Long symmetric cyclic protocol (LS)},
Fig.~\ref{fig:scheme}(c).  This is basically an S-protocol with a
doubled control interval, $\Delta t \mapsto 2\Delta t$.  Equivalently,
note that this scheme corresponds to alternating a $\pi$-pulse with
the identity after every $\Delta t$.  The cycle time becomes
$T_c=4\Delta t$.  For this amount of time, twice as many pulses would
be used by protocols (A, S). Still, in certain cases, the LS-protocol
performs better than the A-protocol (see Sect.~V.D), which motivates
its separate consideration here.

$\bullet$ {\bf Naive random protocol (R)}, Fig.~\ref{fig:scheme}(d).
Random decoupling is no longer cyclic, meaning that the control
propagator does {\em not} necessarily effect a closed path (see
also~\cite{Facchi2004} for a discussion of acyclic deterministic
schemes).  The simplest random protocol in our setting corresponds to
having, at each time $t_j=t_0+j\Delta t$, an equal probability of
rotating or not the qubit that is, at every $t_j$ the control action
has a 50\% chance of being a $\pi$-pulse and a 50\% chance of being
the identity.  In order not to single out the first control slot, it
is convenient to explicitly allow the value $j=0$ (equivalently, to
consider a fictitious pulse $P_0=\openone$ in the A-, S- and
LS-protocols).  For pure phase errors as considered, such a protocol
may be interpreted as a simplified PAREC scheme~\cite{Kern2004}.
While we will mostly focus on this naive choice in our discussion
here, several variants of this protocol may be interesting in
principle, including unbalanced pulse probabilities and/or
correlations between control operations.

$\bullet$ {\bf Hybrid protocol (H)}, Fig.~\ref{fig:scheme}(e).
Interesting control scenarios arise by combining deterministic and
random design.  The simplest option, which we call ``hybrid'' protocol
here, consists of alternating, after every $\Delta t$, a $\pi$-pulse
with a random pulse, instead of the identity as in the LS-protocol.
For our system, in the embedded decoupling language
of~\cite{Kern2005}, this may be thought of as nesting the A- and
R-protocols. In group-theoretic terms, the H-protocol may be
understood as {\em randomization over
cycles}~\cite{Viola2005Random}. A complete asymmetric cycle may be
constructed in two ways, say $A_1$ and $A_2$.  Cycle $A_1$ corresponds
to traversing ${\cal G}$ in the order $(\openone, \sigma_x )$ that is,
free evolution for $\Delta t$; first pulse; free evolution for $\Delta
t$; second pulse -- the cycle being completed right after the second
pulse.  Cycle $A_2$ corresponds to the reverse group path, $(\sigma_x,
\openone)$.
Thus, we have: pulse; free evolution for $\Delta t$; second pulse; and
another free evolution for $\Delta t$ -- the system should be observed
at this moment before any other pulse.  The H-protocol consists of
uniformly picking at random one of the two cycles at every instant
$t_{2j}$, where $j=0,1,\ldots $

\section{Randomized phase refocusing in an isolated qubit}

A single-qubit evolving according to unitary dynamics $(\mu =0$ in
Eq.~(\ref{drift})) provides a pedagogical yet illustrative setting to
study dynamical control. Since the goal here is to refocus the
underlying phase evolution, the analysis of this system provides a
transparent picture for the differences associated with deterministic
and random pulses.  It also simplifies the comprehension of the
results for the more interesting case of a single-qubit interacting
with a decohering semiclassical or quantum environment, where the
control purpose becomes twofold: phase refocusing and decoherence
suppression.

\subsection{Time-independent qubit Hamiltonian}

We begin by considering the standard case of a time-independent target
dynamics, $\omega_0(t) \equiv \omega_0$ for all $t$.  For all the
control protocols illustrated above, the system evolves freely between
pulses, with the propagator
\begin{equation}
U_0(t_{j+1}, t_j ) = 
e^{-i \omega_0(t_{j+1}-t_j) \sigma _z /2} \,,
\label{for_apA}
\end{equation}
whereas, during a pulse, it is only affected by the control
Hamiltonian.  The propagator for an instantaneous pulse applied at
time $t=t_j$ will be indicated by $P_j$. Let
\begin{equation}
\rho(t)=\sum_{\ell, m=0,1} \rho_{\ell m}(t) |\ell\rangle\langle m|
\label{rho}
\end{equation}
denote the qubit density operator in the computational basis
$\{|0\rangle,|1\rangle \}$, with $\sigma_z |0\rangle = |0\rangle$ and
$\sigma_z |1\rangle = -|1\rangle$.  The relevant phase information is
contained in the off-diagonal matrix element $\rho_{01}(t)$.  If
$\rho(t_0)$ is the initial qubit state, the time evolution after $M$
control intervals under either deterministic or 
randomized protocols,
\begin{equation} 
\rho(t_M)=U(t_M,t_0) \rho (t_0) U^{\dagger } (t_M,t_0) \:,
\label{rho_single}
\end{equation}
is dictated by a propagator of the form 
\begin{widetext}
\begin{eqnarray}
&&U(t_M,t_0) = {\cal T}\exp \left\{\hspace*{-1mm}-i 
\int_{t_0}^{t_M} [H_0+ H_c(u)]\, du \right\}
=P_M U_0(t_M,t_{M-1})P_{M-1} U_0(t_{M-1},t_{M-2}) 
\ldots P_1 U_0(t_1,t_0)P_0 \nonumber \\
&& = \underbrace{(P_M P_{M-1} \ldots P_1 P_0)}
(P_{M-1} \ldots P_1 P_0)^{\dagger} 
U_0(t_M,t_{M-1})
(P_{M-1} \ldots P_1 P_0)\ldots 
U_0(t_2,t_1)(P_1 P_0) P_0^{\dagger} U_0(t_1,t_0) P_0\:, 
\label{U_single} \\
&&\hspace{1.5 cm} U_c(t_M) \nonumber
\end{eqnarray}
\end{widetext}
where ${\cal T}$ indicates, as usual, time ordering.

Recall the basic idea of deterministic phase refocusing.  For the
A-protocol, $P_0=\openone$ and $P_j= \exp( -i {\pi} \sigma _x/2 )$,
$j\geq1$ [see Appendix A]. Exact averaging is then ensured after a
single control cycle, thanks to the property
\begin{equation}
P_1^\dagger e^{-i \omega_0\Delta t \sigma _z /2   } P_1
= e^{+i \omega_0 \Delta t \sigma _z /2}
\:.
\label{property}
\end{equation}
Thus, the total phase that the qubit would accumulate in the absence
of control is fully compensated, provided that $N$ complete cycles are
effected (that is, an even number $M=2N$ of spin flips is applied).
The overall evolution implements a stroboscopic no-op gate,
$U(t_M,t_0) = \openone$, $(t_M-t_0)=M\Delta t = NT_c $, as
desired~\cite{remark}.  Notice that the identity operator
is also recovered with the S- and LS-protocols after their
corresponding cycle is completed.

\subsubsection{Control performance in the logical frame}

In preparation to the randomized protocols (R,H), it is instructive to
look at the system dynamics in a different frame. In particular, a
formulation which is inspired by NMR~\cite{ErnstBook} is the so-called
{\em toggling-frame} or {\em logical-frame} picture, which corresponds
to a time-dependent interaction representation with respect to the
applied control Hamiltonian.  Let
\begin{equation}
U_c(t,t_0) = {\cal T} \exp \left\{- i\int _{t_0}^t H_c (u) du 
\right\}
\end{equation}
denote the control propagator associated to $H_c(t)$. Then the
transformed state is defined as
\begin{equation}
\tilde{\rho} (t) = U^{\dagger }_c(t,t_0) \rho  (t) U_c (t,t_0)\:,
\label{rho_logical}
\end{equation}
with tilde indicating henceforth logical-frame quantities.  At the
initial time $t_0$, the two frames coincide and $\tilde{\rho}
(t_0)=\rho (t_0)$.  The evolution operator in the logical frame is
immediately obtained from Eqs. (\ref{rho_single}) and
(\ref{rho_logical}),
\begin{equation}
\tilde{U}(t,t_0)=U^{\dagger }_c(t,t_0)U(t,t_0)\:,
\label{correspondence}
\end{equation}
with 
\begin{equation}
\tilde{U}(t,t_0) ={\cal T}\exp \left \{  -i \int_{t_0}^t 
[U^{\dagger }_c(u) H_0 U_c (u)]\, du
\right\} \:.
\end{equation}
That is, the control field is {\em explicitly removed} from the
effective logical Hamiltonian.  Because, for BB multipulse control,
\begin{equation}
U_c(t_M,t_0)= P_M P_{M-1} \ldots P_1 P_0\:, 
\label{Uc}
\end{equation}
the expression for the logical frame propagator may simply be read off
Eq.~(\ref{U_single}), yielding
\begin{equation}
\tilde{U}(t_M,t_0)={\cal T}\bigg( \prod_{j=0}^{M-1} {\cal P}^{\dagger }_j 
U(t_{j+1}, t_j ) {\cal P}_j \bigg) \:,
\label{U_logical}
\end{equation}
in terms of the composite rotations 
\[ {\cal P}_j=P_j P_{j-1} \ldots P_1 P_0\:, 
\hspace{5mm}j=0,\ldots , M-1\:. \]

For cyclic protocols, $U_c(t_M,t_0)= \exp ( -i M {\pi}
\sigma_x/2)=\openone$ ($M$ even) that is, the logical and physical
frames overlap stroboscopically in time.  Thus,
$\tilde{U}(t_M,t_0)=\openone $ and {\em phase refocusing in the
logical frame is equivalent to phase refocusing in the physical
frame.}

Now consider the evolution under the randomized protocols.  The first
pulse occurs at $t_0$, so after a time interval $t_M-t_0$ has elapsed,
$M+1$ pulses have been applied.  Since the final goal is to compare
random with cyclic controls, we shall take $M$ {\em even} henceforth.
At time $t=t_M$, population inversion may have happened in general in
the physical frame.  This makes it both convenient and natural to
consider the logical frame, where inversion does {\em not} happen, as
the {\em primary} frame for control design. The evolution operator in
this frame may be expressed, using Eq.~(\ref{U_logical}), as
\begin{equation}
\tilde{U}(t_M,t_0) = \exp \bigg\{ -i \frac{\omega_0\Delta t}{2} \sigma_z 
\sum_{j=0}^{M-1} \chi _j \bigg\}\:,
\label{U_logical_chi}
\end{equation}
where
\begin{equation}
\chi _j = (-1)^{\lambda _0 + \lambda _1 + \ldots +
\lambda _j }\:, \hspace{5mm}j=0,\ldots , M-1\:,
\label{chi}
\end{equation}
is a Bernoulli random variable which accounts for the history of spin
flips up to $t_j$ in a given realization.  For each $m=1,\ldots,j$, if
a spin flip occurs at time $t_m$, then $\lambda_m=1$ and $P_m= -i
\sigma _x$, otherwise $\lambda _m=0$ and $P_m=\openone$. Equivalently,
$\chi_j$ will take the values $+1$ or $-1$ with equal probability,
depending on whether the composite pulse ${\cal P}_j$ is the identity
or a $\pi$-pulse.

Let $k$ be an index labelling different control realizations. For a
fixed $k$, the qubit coherence in the logical frame is given by
\begin{equation}
\tilde{\rho}_{01}^{(k)}(t_M)= e^{ -i \omega_0 \Delta t
\sum_{j=0}^{M-1} \chi _j^{(k)} } \rho_{01}(t_0)\:.
\label{singlereal}
\end{equation}
This expression provides the starting point for analyzing control
performance. For the A-protocol, the only possible realization has
$\chi_j = (-1)^j$ and leads to the trivial result
${\tilde{\rho}_{01}}(t_M)= \rho_{01}(t_0)$.  For the R-protocol,
realizations corresponding to different strings of $\lambda$'s filling
up $M$ places give, in general, different phases and an ensemble
average should be considered.  If the statistical ensemble is large
enough, the {\em average performance} may be approximated by the {\em
expected performance}, which is obtained by averaging over {\em all}
possible control realizations and will be denoted by $\mathbb{E}(\:)$.
The calculation of the expectation value is straightforward in the
{\em unbiased} setting considered here.  Since, for each realization,
$\chi_j = +1$ or $-1$ independently of the value of its predecessor
$\chi_{j-1}$, 
the following expression is found: 
\begin{equation}
\mathbb{E}\Big(\tilde{\rho}_{01}(t_M)\Big)=
{\rho_{01}(t_0)} [\cos (\omega_0 \Delta t)]^M \,.
\label{E_logical}
\end{equation}

Several remarks are in order.  Under random pulses, the phase
accumulated during the interval $t_M-t_0=M\Delta t$ is, on average,
completely removed, regardless of the $\Delta t$ value.  An important
distinction with respect to the deterministic controls, however, is
that now the different phase factors carried by each stochastic
evolution may interfere among themselves, causing the ensemble average
to introduce an effective {\em phase damping}.  In general, let us
write the ensemble expectation in the form
\begin{equation}
\frac{\mathbb{E}\Big(\tilde{\rho}_{01}(t_M)\Big)}
{\rho_{01}(t_0)}=
e^{i \phi_{*} (t_M,t_0)} e^{-\Gamma_{*} (t_M,t_0)},
\label{deph_expform}
\end{equation}
for real functions $\phi_{*} (t), \Gamma_{*} (t)$~\cite{note*}. Here,
$\phi_{*} (t_M,t_0)=0$, whereas $\Gamma_{*} (t_M,t_0)=-M\ln [\cos
(\omega_0 \Delta t)]$.  Complete dephasing occurs when $\omega_0
\Delta t =l' \pi/2$, with $l'$ odd, while for $\omega_0 \Delta t =
l\pi $, with $l \in {\mathbb Z}$, $\Gamma_{*} (t)=0$.  Whenever exact
knowledge of the frequency $\omega_0$ and precise control over the
time interval $\Delta t$ are available, the R-protocol can be made to
achieve exact averaging, like the A-protocol, under the additional
{\em synchronization condition} that
\[ \Delta t = l\pi /\omega_0\:, \;\;\;  l \in {\mathbb Z}\:.\]  

In situations where such a synchronization is not easily accessible,
one may still look for a general condition under which the R-protocol
avoids ensemble dephasing.  Taking a Taylor expansion of
Eq.~(\ref{E_logical}) yields
\begin{equation}
\omega_0^2 (t_M-t_0) \Delta t=\omega_0^2 (t_M-t_0)^2/M\ll 1\:.
\label{first_scale}
\end{equation}
In principle, this requirement may be fulfilled by making $t_M$ and/or
$\Delta t$ sufficiently small.  Interestingly, the condition of
Eq.~(\ref{first_scale}) is directly related to the bound obtained in
Theorem 1 of \cite{Viola2005Random} for the worst-case pure-state
error probability, defined by
\begin{equation}
\varepsilon_{t} =\max_{|\psi\rangle}
\{ \varepsilon_{t}(|\psi\rangle) \} 
=1-\min_{|\psi\rangle}
\mathbb{E} \big( {\rm Tr}
(\rho(t_0) \tilde{\rho} (t)) \big) \:,
\end{equation}
where the latter term is the usual input-output state
fidelity~\cite{NielsenBook}. 
In the limit where $||H_0(t)||_2^2 \,t \Delta t\ll1$,
where $||A||_2=\max |{\rm eig} (A)|$, $\forall A=A^{\dagger}$,
Theorem 1 implies
\begin{equation}
\varepsilon_{t}=
{\cal O} \Big( ||H_0(t)||_2^2 \, t \,\Delta t \Big) \:.
\label{error_bound_single}
\end{equation} 
On the other hand, using Eq.~(\ref{E_logical}) we obtain
\begin{eqnarray*}
\varepsilon_{t_M}(|\psi\rangle)=
2 \bigg(\hspace*{-0.5mm}\rho_{00}(t_0) \rho_{11}(t_0) -
|\rho_{01}(t_0) |^2 [\cos (\omega_0 \Delta t)]^M \hspace*{-0.5mm}\bigg) \,.
\end{eqnarray*}
For $\omega_0^2 (t_M-t_0) \Delta t\ll 1$, the above expression gives
\begin{eqnarray}
&&\varepsilon_{t_M}(|\psi\rangle)\approx|\rho_{01}(t_0)|^2\,
\omega_0^2 (t_M-t_0) \Delta t \:, \\
&&\varepsilon_{t_M}=
{\cal O} \Big( \omega_0^2 (t_M-t_0) \Delta t \Big) \:,
\end{eqnarray}
which makes the connection with Eq.~(\ref{first_scale}) manifest.

It remains to discuss the performance of the H-protocol.  The freedom
of not always effecting a spin flip after every $\Delta t$, which is
one of the appealing features of the R-protocol, is still partially
present here. On the other hand, since a spin flip does occur at every
$t_m$ with $m$ odd, which leads to $\chi_j=-\chi_{j-1}$ for $j$ odd,
any realization of this protocol completely refocuses the qubit [see
Eq.~(\ref{singlereal})], so $\phi_{*} (t_M,t_0)=0$ and $\Gamma_{*}
(t_M,t_0)=0$.  Accordingly, {\em in the logical frame, the H-protocol
is optimal}, combining the absence of phase damping of cyclic schemes
with the flexibility of random pulses.

\subsubsection{Ensemble averages: General remarks}

In practice, we deal with the average performance of a statistical
ensemble of size $K$.  To evaluate the sample size that guarantees a
desired margin of error $\delta $ \cite{Mansfield_statistics}, we
invoke the central limit theorem. Because different realizations are
independent, the latter ensures that the average performance is
distributed normally with a mean value equal to the expected
performance and standard deviation given by $\sigma /\sqrt{K}$, where
$\sigma $ is the standard deviation for all realizations. Thus, if we
want, with probability $(1-\epsilon )$, that the average performance
differs from the expected performance by no more than $\delta $, the
sample size must be at least as large as
\begin{equation}
K_{{\rm min}}=\left(\frac{z_{\epsilon/2} \sigma }{\delta}\right)^2 = 
{\cal O} \left( \frac{\sigma^2}{\delta^2} \right),
\end{equation}
where $z_{\epsilon/2}$ is the value of the standard normal variable
which has a probability $\epsilon/2$ of being exceeded.  Taking a Taylor
expansion of Eqs.~(\ref{singlereal}) and (\ref{E_logical}), we can
show that
$$\sigma ={\cal O}(\omega_0\sqrt{(t_M-t_0) \Delta t}) \hspace{0.4 cm}
{\rm for} \hspace{0.4 cm} \omega_0^2 (t_M-t_0) \Delta t\ll1 \:. $$
Thus, the number of realizations required to ensure a specified degree
of precision decreases as $\Delta t$.

It is interesting to observe that the ensemble average may be
interpreted as effecting a quantum operation,
\[ \mathbb{E}\Big(\tilde{\rho}_{01}(t_M)\Big)=
\sum_{k} \frac{\tilde{U}^{(k)}}{\sqrt{2^M}} \tilde{\rho}(t_0) 
\frac{\tilde{U}^{(k)\dagger}}{\sqrt{2^M}} \]
with
\[ \tilde{U}^{(k)}(t_M,t_0)=\alpha^{(k)} \openone + \beta^{(k)}
\sigma_z\:, \hspace{5mm} \sum_k \tilde{U}^{(k)\,\dagger}
\tilde{U}^{(k)} =\openone\:, \] 
and random coefficients $\alpha^{(k)},
\beta^{(k)}$ which may be derived from Eq.~(\ref{U_logical_chi}).

\subsubsection{Control performances in the physical frame}

Finally, it is important to compare the average coherence element in
the logical and physical frames.  Dephasing is a more delicate issue
in the Schr\"odinger picture, because spin population is not
necessarily conserved and $\rho_{01}(t_M)$ may be related to
$\rho_{01}(t_0)$ or to $\rho_{10}(t_0)$, depending on how many
$\pi$-pulses occur.  If, after an interval $t_M-t_0$, an even number
of spin-flips have happened, we recover $U_c(t_M,t_0)=\openone$ as in
the cyclic case, but an odd number of flips leads instead to
$U_c(t_M,t_0) = \pm i \sigma _x$. By recalling
Eq.~(\ref{correspondence}), for randomized schemes we find
\begin{equation}
\mathbb{E}(\rho_{01}(t_M))=
\frac{\rho_{01}(t_0) 
+\rho_{10}(t_0) }{2} \, e^{-\Gamma_{*}(t_M,t_0)}\:,
\label{E_physical}
\end{equation}
where $\Gamma_{*} (t_M,t_0)=0$ for the H-protocol.  Thus, the
agreement between the expected results in the two frames depends on
the initial qubit state. Results are identical if $\rho_{01}(t_0)$ is
real, but differ otherwise. The worst scenario occurs if
$\rho_{01}(t_0)$ is purely imaginary, as the average in the physical
frame vanishes.  This reflects the fact that the net evolution may be
represented by a quantum operation which flips the state of the qubit
with 50\% probability, and leaves it alone otherwise.  Clearly,
knowledge of the control history allows the system to be
deterministically returned in the physical frame for any realization,
if desired. That is, having a classical register that records the
total number of spin flips may be used to select realizations that
guarantee a good performance of random pulses also in the physical
frame for any initial state. For example, if only realizations with an
even number of spin-flips are selected, the results in both frames are
equal, $\mathbb{E}(\rho_{01}(t_M)|U_c(t_M,t_0=\openone)=
\mathbb{E}(\tilde{\rho}_{01}(t_M))$, as desired.

{\em To summarize:} In the logical frame, refocusing the unwanted
phase evolution is possible with any of the protocols we considered.
The R-protocol, however, introduces an average ensemble dephasing,
which may only be prevented by precisely tuning $\Delta t =
l\pi/\omega_0 $, with $l \in {\mathbb Z}$, or by assuring that $
\Delta t\ll 1/[\omega_0^2 (t_M-t_0)]$.  This implies the appearance of
a time scale requirement which is not present when dealing with
deterministic controls, nor with the H-protocol.  In the physical
frame, state-independent conclusions regarding the system behavior may
be drawn conditionally to specific subsets of control realizations.
Overall, the H-protocol emerges as an alternative of intermediate
performance, which partially combines advantages from determinism and
randomness.

\subsection{Time-dependent qubit Hamiltonian}

We now consider the more interesting case where the qubit frequency is
time dependent, $\omega_0(t)=\omega_0 + \delta \omega_0(t)$, $\delta
\omega_0(t)\equiv \omega_0 G(t)$ being a deterministic (but
potentially unknown) function.  This could result, for example, from
uncontrolled drifts in the experimental apparatus.

While all protocols become essentially equivalent in the limit
$M\rightarrow \infty$, searching for the best protocol becomes
meaningful in practical situations where pulsing rates are necessarily
finite.  Under these conditions, the deterministic protocols described
so far will no longer be able, in general, to completely refocus the
qubit. This would require a very specific sequence of spin flips for
each particular function $\delta \omega_0(t)$, which would be hard to
construct under limited knowledge about the latter.  On the other
hand, the average over random realizations does remove the phase
accumulated for {\em any} function $\delta \omega_0(t)$, making
randomized protocols ideal choices for phase refocusing. As a drawback,
however, ensemble dephasing may be introduced.  Thus, the selection 
of a given protocol will be ultimately dictated by the resulting
tradeoffs.

The propagator in the logical frame now reads
\begin{equation}
\label{U_time_isol}
\tilde{U}(t_M,t_0) =\exp \bigg[ -i \frac{\omega_0}{2} \sigma_z 
\sum_{j=0}^{M-1}  \chi_j
\int_{t_j}^{t_{j+1}} \hspace{-0.3cm} (1 + G(u)) du 
\bigg]  \:,
\end{equation}
which reduces to Eq.~(\ref{U_logical_chi}) when $G(t)=0$.

Some assumptions on both the amplitude and frequency behavior of
$G(t)$ are needed in order to draw some general qualitative
conclusions.  First, if $|G(t)|\ll 1$, the analysis developed in the
previous section will still approximately hold.  In the spirit of
regarding $\omega_0$ as a central frequency, we will also discard the
limit $|G(t)|\gg 1$, and restrict our analysis to cases where $\max_t
\,|G(t)| \sim 1$.  If $G(t)$ is dominated by frequency components
which are very fast compared to $\tau_0=\omega_0^{-1}$, the effect of
$G(t)$ may effectively self-average out over a time interval of the
order or longer than $\omega_0^{-1}$.  In the opposite limit, where
the time dependence of $G(t)$ is significantly slower than $\omega_0$,
deterministic controls are expected to be most efficient in refocusing
the qubit, improving steadily as $\Delta t$ decreases.  In
intermediate situations, however, the deterministic performance may
become unexpectedly poor for certain, in principle, unknown values of
$\Delta t$.  These features may be illustrated with a simple periodic
dependence.  Suppose, for example, that $G(t)=\sin (p\, \omega_0 t)$,
and $p\in \mathbb{R}$. For a fixed time interval $t_f-t_0 < \pi/(p
\,\omega_0)$, a significant reduction of the accumulated phase is
already possible with few deterministic pulses.  However, care must be
taken to avoid unintended ``resonances'' between the natural and the
induced sign change.  For the A-protocol, this effect is worst at
$\Delta t=\pi/(p \,\omega_0)$, in which case the control pulses
exactly occur at the moment the function changes sign itself, hence
precluding any cancellation of $G(t)$.

With the R-protocol, ensemble dephasing becomes the downside
to face. The ensemble average now becomes
\begin{equation}
e^{-\Gamma_{*} (t_M,t_0)}=\prod _{j=0}^{M-1} \cos \bigg\{ \omega_0 \bigg[
\Delta t + \int_{t_j}^{t_{j+1}} G(u) du 
\bigg] \bigg\}\,. 
\end{equation}
In the absence of time dependence, phase damping is minimized as long
as Eq.~(\ref{first_scale}) holds.  Under the above
assumptions on $G(t)$, the condition remains essentially unchanged, in
agreement with the fact that the accuracy of random averaging only
depends on $|| H_0 (t) ||_2$~\cite{Viola2005Random}.

Refocusing is also totally achieved with the H-protocol. However,
unlike in the case of the R-protocol, the ensemble average no longer
depends on the time independent part of the Hamiltonian, but only on
the function $G(t)$, making the identification of precise requirements
on $\Delta t$ harder in the absence of detailed information on the
latter.  We have
\begin{equation}
e^{-\Gamma_{*} (t_M,t_0)}= 
\hspace*{-2.5mm}\prod _{j=0,2,4,\ldots}^{M-2} 
\hspace*{-3.5mm}\cos \bigg\{ \omega_0 \bigg[\bigg(
\int_{t_j}^{t_{j+1}} \hspace*{-1.5mm} -  \int_{t_{j+1}}^{t_{j+2}}\bigg)
G(u) du \bigg] \bigg\}\,.
\end{equation}

\begin{figure}[htb]
\includegraphics[width=3.in]{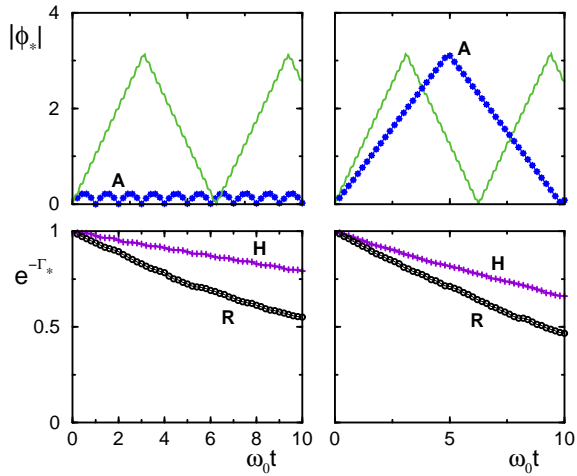}
\caption{(color online) Accumulated phase (upper panels) and dephasing
rate (lower panels) in the absence of control [solid (green) line],
and under the A- [(blue) stars], R- [(black) circles], and H-
[(purple) plus] protocols in the logical frame, for
$G(t)=\sin(p\,\omega_0 t)$ and $\Delta t=1/(10\omega_0)$.  Left
panels: $p=20\sqrt{2}$; right panels: $p=10\pi$. Average taken over
$10^3$ realizations.  In this and all simulations that will 
follow, we set $t_0=0$.   }
\label{fig:fixedM}
\end{figure}

Fig.~\ref{fig:fixedM} illustrates the points discussed so far.  The
sinusoidal example is considered, and we contrast the two aspects to
be examined: the top panels show the phase magnitude
$|\phi_{*}(t_M,t_0)|$, which is optimally eliminated with random
pulses, while the bottom ones give the dephasing rate $e^{-\Gamma_{*}
(t_M,t_0)}$, which is inexistent for deterministic controls.  The
interval between pulses is fixed, $\Delta t=1/(10\omega_0)$, and the
protocols are compared for two arbitrary, but relatively close values
of the oscillation frequency rate: $p=20\sqrt{2}$ and $p=10\pi$. The
deterministic control is very sensitive to slight changes of the drift
and at certain instants may behave worse than if pulses were
completely avoided. Similarly, the H-protocol, even though more
effective than the R-protocol in this example, also suffers from
uncertainties related to $G(t)$.  On the contrary, deviations in the
performance of the R-protocol are practically unnoticeable, making it
{\em more robust against variations in the system parameters}.

\begin{figure}[htb]
\includegraphics[width=3.in]{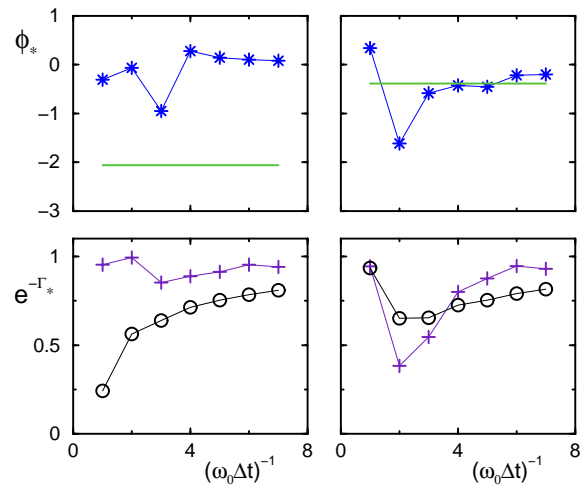}
\caption{(color online) Accumulated phase (upper panels) and dephasing
rate (lower panels) in the absence of control [solid (green) line],
and under the A- [(blue) stars], R- [(black) circles], and H-
[(purple) plus] protocols in the logical frame.  The time interval
considered is $t_f=2/\omega_0$; $(\omega _0 \Delta t)^{-1}=M/2$; and
$G(t)=\sin (p\,\omega_0 t)$, where $p=10$. The drift in the right
panels includes $D(t)=(-1)^{\lfloor 10 \omega_0 t/3\rfloor }$.
Average taken over all possible realizations. }
\label{fig:w_sin}
\end{figure}

As a further illustrative example, we consider in Fig.~\ref{fig:w_sin}
the following time dependence for the qubit:
\begin{equation}
\omega_0(t)=\omega_0 [1+G(t)] D(t).
\end{equation}
The left panels have, as before, $D(t)=1$, while for the right
panels
\begin{equation}
D(t)=(-1)^{\lfloor 10 \omega_0 t/3\rfloor }.
\end{equation}
A fixed time $t_f=2/\omega_0$ is now divided into an increasing number
$M$ of intervals $\Delta t$.  Here, selecting the most appropriate
protocol depends on our priorities concerning refocusing and
preservation of coherence.  We may, however, as the right upper panel
indicates, encounter {\em adversarial} situations where the time
dependence of the qubit frequency is such that not acting on the
system is comparatively better than using the A-protocol.  Clearly,
depending on the underlying time dependence and the pulse separation,
such poor performances are also expected to occur with other
deterministic protocols.  In addition, notice that, consistent with
its hybrid nature, the H-protocol may perform worse for
values of $\Delta t$ where the deterministic control becomes 
inefficient (compare right upper and lower panels).  In similar
situations, from the point of view of its enhanced stability, {\em the
R-protocol turns out to be the method of choice}.

{\em To summarize:} An isolated qubit with time-dependent parameters
provides the simplest setting where advantages of randomization begin
to be apparent, in terms of enhanced stability against parameter
variations.  On average, phase is fully compensated, and ensemble
dephasing may be kept very small for sufficiently fast control.
Similar features will appear for a qubit interacting with a
time-varying classical or quantum environment, as we shall see in
Sects.~IV.C and V.E.

\section{Randomized control of decoherence from a semiclassical 
environment}

Qubit coherence is limited by the unavoidable influence of noise
sources.  Within a semiclassical treatment, which provides an accurate
description of decoherence dynamics whenever back-action effects from
the system into the environment can be neglected, noise is modeled in
terms of a classical stochastic process, effectively resulting in
randomly time-dependent systems.  Typically, external noise sources,
which in a fully quantum description are well modeled by a {\em
continuum} of harmonic modes (see Sect.~V), are represented by a
Gaussian process.  Here, we focus on {\em localized} noise sources,
which may be intrinsic to the physical device realizing the qubit --
notably, localized traps or background charges, leading to a quantum
{\em discrete} environment.  In this case, non-Gaussian features
become important, and are more accurately represented in terms of
noise resulting from a single or a collection of classical bistable
fluctuators -- leading to so-called random telegraph noise (RTN) or
$1/f$-noise, respectively.  Beside being widely encountered in a
variety of different physical
phenomena~\cite{Weissman1988,Press1978,Voss1992}, such noise
mechanisms play a dominant role in superconducting Josephson-junction
based implementations of quantum
computers~\cite{Makhlin2001,Paladino2002,Itakura2003,Galperin2003}.

Recently, it has been shown that RTN and $1/f$-noise may be
significantly reduced by applying cyclic sequences of BB
pulses~\cite{Shiokawa2002,Gutmann2003,Falci2004,Faoro2004,Gutmann2004}.
We now extend the analysis to randomized control.  As it turns out,
random decoupling is indeed viable and sometimes more stable than
purely deterministic protocols.  While a detailed analysis of
randomized control of genuine $1/f$ noise would be interesting on its
own, we begin here with the case of a single fluctuator.  This
provides an accurate approximation for mesoscopic devices where noise
is dominated by a few fluctuators spatially close to the
system~\cite{Rogers1984,Wakai1987,Gutmann2003,Galperin2003}.  Let the
time-dependent Hamiltonian describing the noisy qubit be given by
Eqs.~(\ref{drift0})-(\ref{drift}), where $\mu=0$ and
\begin{equation}
\delta \omega_0 (t) ={\rm RTN}(t)
\label{Ham_tele}
\end{equation}
characterizes the stochastic process, randomly switching between two
values $\pm v/2$, $v>0$.  We shall in fact consider a {\em semi}random
telegraph noise, that is, we assume that the fluctuator initial state
is always $+v/2$.  The switching rate from $\pm v/2$ to $\mp v/2$ is
denoted by $\gamma _{\mp}$, with $\gamma_+ +\gamma_-=\gamma $.  We 
shall also assume for simplicity that $\gamma_+ =\gamma_-$, 
corresponding to a symmetrical process.
The number of switching events $n(t,0)$ in a given
time interval $t$ is Poisson-distributed as
\[ P\big\{n(t,0)=k\big\}=\frac{1}{k!}
\left(\frac{\gamma t}{2}\right)^k e^{-\gamma t/2}\:. \] 

\begin{figure}[tb]
\includegraphics[width=2.4in]{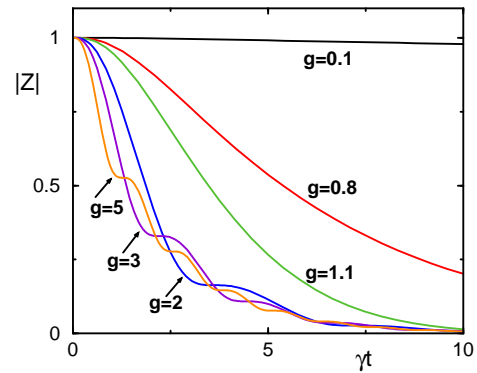}
\caption{(color online).  Decoherence rate $|Z(t)|=e^{-\Gamma (t)}$
from a symmetrical bistable fluctuator.  Several values of
$g=v/\gamma$ are considered, resulting from changing the coupling
strength $v$ at fixed switching rate $\gamma=1$ a.u.}
\label{fig:tele_nopulse}
\end{figure}

Semiclassically, dephasing results from the ensemble average over
different noise realizations.  This leads to the decay of the average
of the coherence element,
\begin{eqnarray}
&&\frac{\langle \rho_{01}(t)\rangle}{\rho_{01}(t_0)}
=e^{-i \omega_0 (t - t_0)} Z(t,t_0) , \nonumber \\
&&Z(t,t_0)=e^{i \delta (t,t_0)} e^{-\Gamma (t,t_0)}\:.
\end{eqnarray}
Here, the average over RTN realizations is represented by $\langle\;
\rangle$ and should be distinguished from the average over control
realizations, which, as before, is denoted by $\mathbb{E}$.  The
dephasing factor $\Gamma (t,t_0)$ and the phase $\delta (t,t_0)$ have
distinctive properties depending on the ratio $g=v/\gamma $, where
$g<1$ ($g>1$) corresponds to a fast (slow) fluctuator.  Given the
initial condition for the fluctuator ${\cal E}_{p_0}$, where ${\cal
E}_{p_0}=\pm 1$ stands for the fluctuator initially in state $\pm
v/2$, $Z(t,t_0)$ may be calculated as~\cite{Paladino2002},
\begin{equation}
Z(t,t_0) = C e^{-\frac{\gamma}{2} (1-\alpha)(t-t_0)}
+(1- C) e^{-\frac{\gamma}{2} (1+\alpha)(t-t_0)},
\end{equation}
where
\begin{equation}
\alpha = \sqrt{1-g^2 + 2 i g {\cal E}_p} \hspace{0.5cm}
C=(1+\alpha-ig{\cal E}_{p_0})/(2\alpha )\:,
\end{equation}
and ${\cal E}_p=(\gamma_- - \gamma_+)/\gamma $ is the equilibrium
population difference.  Note that for a symmetrical telegraph process,
the only difference between the results for a fluctuator initially in
state $+v/2$ or $-v/2$ is a sign in the above phase $\delta $.  The
decoherence rate for a slow fluctuator is much more significant than
for a fast fluctuator. This has been discussed in detail elsewhere
\cite{Paladino2002}, and has been reproduced for later comparison with
the controlled case in Fig.~\ref{fig:tele_nopulse}, where several
values of $g$ are considered.  A fast fluctuator behaves equivalently
to an appropriate environment of harmonic oscillators and noise
effects are smaller for smaller values of $g$, whereas for a slow
fluctuator the decoherence function saturates and becomes $\sim \gamma
(t-t_0)$.

\subsection{Deterministic and randomized 
controls in the interaction picture}

Here we compare the reduction of RTN under the action of the A-, H-,
and R-protocols.  In order to isolate the effects of the noise, it is
convenient to first carry out the analysis in the interaction picture
which removes the free dynamics $\omega_0 \sigma_z/2$. The density
operator becomes
\begin{equation}
\rho^I(t)=U^I(t,t_0) \rho (t_0) U^{I\dagger} (t,t_0)\:,
\end{equation}
where $U^I(t,t_0)=\exp \left(i\omega_0 t\sigma_z/2\right)$ and the
superscript $I$ will refer to the interaction picture henceforth.  The
free propagator between pulses is now
\begin{equation}
U^I(t_{j+1},t_j) = {\cal T} \exp \left\{-i\int_{t_j}^{t_{j+1}} 
H_0^I (u) du\right\}\:,
\label{Uclas_IP}
\end{equation}
with $H_0^I(t) = \delta \omega_0 (t)\sigma_z/2$; while at $t_j$, we
have [see Appendix A]
\begin{equation}
P^I_j=\exp \left[i\frac{\omega_0 t_j}{2}\sigma _z \right]
\exp \left[-i\lambda _j \frac{\pi}{2} \sigma _x\right]
\exp \left[ -i\frac{\omega_0 t_j}{2}\sigma _z \right]\:.
\label{Pclas_IP}
\end{equation}

A second canonical transformation into the logical frame is also
considered, so that (as before) realizations with an even or an odd
number of total spin flips are treated on an equal footing.  We will
refer to the combination of the two transformations as the logical-IP
frame.  Similarly to Eq.~(\ref{correspondence}), the interaction and
the logical-IP frame propagators are related as
\begin{equation}
\tilde{U}^I(t,t_0)=U_c^{\dagger I}(t,t_0) U^I(t,t_0).
\label{relation}
\end{equation}
This leads to the following propagators at $t_M$,
\begin{eqnarray}
&&U_c^I(t_M,t_0)={\cal T} \bigg( \prod _{j=0}^{M} P^I_j \bigg)\:,
\nonumber \\
&&\tilde{U}^I(t_M,t_0)={\cal T}\bigg( \prod_{j=0}^{M-1} 
{\cal P}^{\dagger I}_j 
U^I(t_{j+1}, t_j ) {\cal P}^I_j \bigg) \:, \nonumber
\end{eqnarray}
where
\[ {\cal P}^I_j = P^I_j P^I_{j-1} \ldots P^I_2 P^I_1 P^I_0\:, 
\hspace{5mm}j=0,\ldots , M-1 \]

Our goal is to compute the ratio
\begin{eqnarray}
{\rm F}(t_M,t_0)&=&\frac{\mathbb{E} \big(\langle 
\tilde{\rho}^I_{01}(t_M)\rangle\big)}
{\rho_{01}(t_0)}
\approx \frac{\langle \mathbb{E} (\tilde{\rho}^I_{01}(t_M))\rangle}
{\rho_{01}(t_0)}
\nonumber \\
&=&
\mathbb{E}\big(e^{i \delta ^{(k)}(t_M,t_0)} 
e^{-\Gamma ^{(k)} (t_M,t_0)}\big) \:,
\end{eqnarray}
where $k$ labels, as before, different control realizations.  Note
that interchanging the order of the averages does {\em not} modify the
results if all pulse realizations are considered and the number of RTN
realizations is large enough.  With $10^5$ switch realizations no
significant variations were found by interchanging the averages.

\begin{figure}[tb]
\includegraphics[width=3.4in]{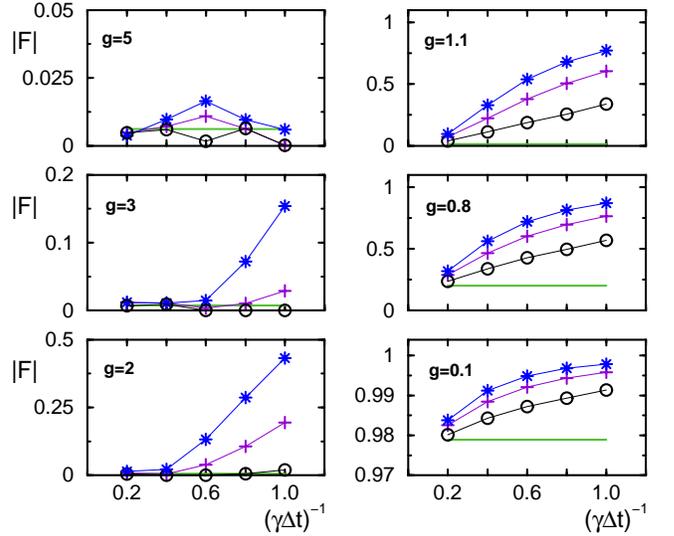}
\caption{(color online) Decoherence rate from a symmetrical bistable
fluctuator with $g=5,3,2,1.1,0.8,0.1$.  (Green) Solid lines represent
the analytical results from~\cite{Paladino2002} in the absence of
control.  (Blue) stars: A-protocol; (black) circles: R-protocol; and
(purple) plus: H-protocol.  Averages are taken over $10^5$ RTN
realizations and all possible $2^M$ pulse realizations.  The time
interval considered is $t_f=10/\gamma$, thus $(\gamma \Delta
t)^{-1}=M/10$. }
\label{fig:tele_dephas}
\end{figure}

The decoherence rate $|{\rm F}(t_M,t_0)|$ for the three selected
protocols is shown in Fig.~\ref{fig:tele_dephas}, where a time
$t_f=10/\gamma$ was fixed and divided into an increasing number $M$ of
intervals $\Delta t$.  The left panels are obtained for three slow
fluctuators, $g=5,3,2$, and the right panels for $g=1.1,0.8,0.1$.
These are the six different noise regimes considered in
Ref.~\cite{Falci2004}, where the A-protocol was studied. The authors
concluded that once $\Delta t\ll1/\gamma$, $\Gamma (t_M,t_0)$ scales
with $g^2$, while for $\Delta t\gtrsim1/\gamma$, BB pulses are still
capable of partially reducing noise due to a fast fluctuator, but are
mostly inefficient against slow fluctuators.  Here, we verified that
among all possible realizations of pulses separated by the same
interval $\Delta t$, the realization corresponding to the A-protocol
yields the largest value of $|{\rm F}(t_M,t_0)|$, whereas absence of
pulses gives, as expected, the smallest value.  This justifies why, in
terms of average performance for finite $\Delta t$, we have in
decreasing order: A-, H-, and R-protocol; while for $M\rightarrow
\infty$, different protocols are expected to become equivalent.

\begin{figure}[t]
\includegraphics[width=3.4in,height=3in]{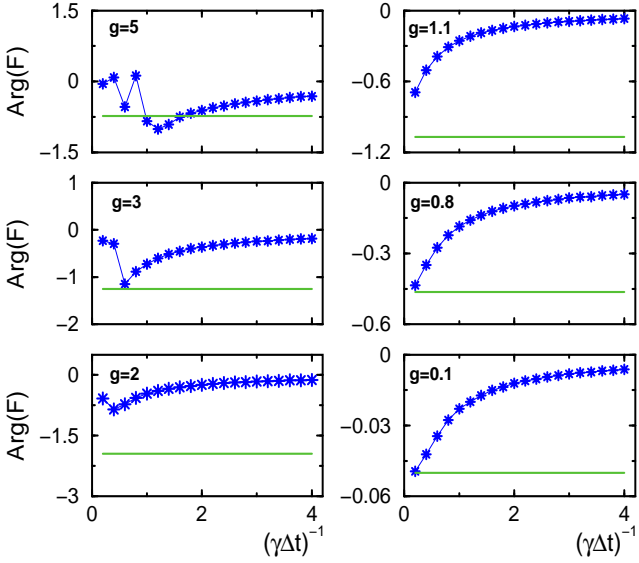}
\caption{(color online) Phase offset from a symmetrical bistable
fluctuator with $g=5,3,2,1.1,0.8,0.1$.  (Green) Solid line: Analytical
results in the absence of control pulses; (blue) stars: A-protocol;
both R- and H-protocols have phase equal to zero.  The time interval
considered is $t_f=10/\gamma$, so $(\gamma \Delta t)^{-1}=M/10$.
Averages computed as in Fig.~\ref{fig:tele_dephas}. }
\label{fig:tele_phase}
\end{figure}

In terms of refocusing the unwanted phase evolution, the above
randomized protocols are optimal, since ${\rm Arg}(\langle \mathbb{E}
(\tilde{\rho}^I_{01}(t_M))\rangle / \rho_{01}(t_0))=0$, while the
phase magnitude $\delta(t_M,t_0)$ for the A-protocol is eventually
compensated as $M$ increases.  This is shown in
Fig.~\ref{fig:tele_phase}. Notice also that the absolute phase is very
small for fast fluctuators.

\begin{figure}[t]
\includegraphics[width=3.4in,height=1.95in]{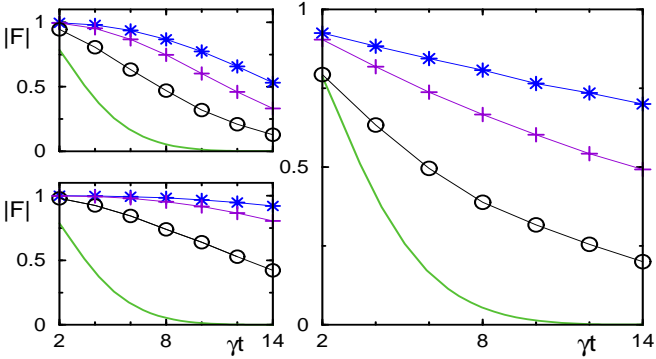}
\caption{(color online) Decoherence rate from a symmetrical bistable
fluctuator with $g=1.1$.  Left panels: $M=10$ (top), $M=30$ (bottom).
Right panel: $\Delta t=1/\gamma$.  (Green) Solid line: Analytical
results in the absence of control.  (Blue) stars: A-protocol; (black)
circles: R-protocol; and (purple) plus: H-protocol.  Averages are
taken over $10^4$ RTN realizations and $10^3$ pulse realizations.}
\label{fig:fixed}
\end{figure}

Instead of fixing a time $t_f$, an alternative picture of the
performance of different protocols may also be obtained by fixing the
number of intervals $M$, as in Figs.~\ref{fig:fixed}~(left panels). As
expected, larger $M$ leads to coherence preservation for longer
times. Still another option is to fix the interval between pulses
$\Delta t$, as in Fig.~\ref{fig:fixed}~(right panel).  As before, the
A-protocol shows the best performance, followed by the H- and
R-protocols.

\subsection{Randomized control in the physical frame}

If the interaction picture is not taken into account, complete
refocusing is again guaranteed, on average, when either the R- or
H-protocols are used.  However, for the R-protocol, the qubit frequency
plays a delicate role in the resulting dephasing process.  We now have
\begin{eqnarray}
&&
\mathbb{E}(\langle \tilde{\rho}_{01}(t_M)\rangle)=  \rho_{01}(t_0)
\cdot \\ 
&&  \mathbb{E}( e^{-i \omega_0 \Delta t
\sum_{j=0}^{M-1} \chi _j^{(k)}  + i \delta ^{(k)}(t_M,t_0) } 
e^{-\Gamma ^{(k)}(t_M,t_0)})\:,\nonumber
\end{eqnarray}
which may be further simplified as follows.  Among the $2^{M}$
pulse realizations existing in the logical frame, there are pairs, say
corresponding to labels $k$ and $k'$, where
$\lambda_0^{(k)}=1$ and $\lambda_0^{(k')}=0$, while
$\lambda_j^{(k)}=\lambda_j^{(k')}$ with $1\leq j\leq M-1$, which
leads to $\sum_{j=1}^{M-1} \chi
_j^{(k)}=-\sum_{j=1}^{M-1} \chi _j^{(k')}$.  Besides, since we are
considering a semirandom telegraph noise, a pulse at $t_0$ is
equivalent to switching the fluctuator from the initial state $+v/2$
to $-v/2$, whose net effect is simply a change in the sign of the
phase $\delta (t_M,t_0)$.  Therefore, we may write
\begin{eqnarray}
&& \mathbb{E}(\langle \tilde{\rho}_{01}(t_M)\rangle)=
\frac{\rho_{01}(t_0)}{2^{M-1}}\cdot \\
&& \sum_{k=1}^{2^{M-1}}\left\{ 
\cos \left[\Xi ^{(k)}_{M-1} \omega_0 \Delta t - \delta ^{(k)}(t_M,t_0)
\right] 
e^{-\Gamma^{(k)}(t_M,t_0)} \right\} \:, \nonumber
\end{eqnarray}
where 
\begin{equation}
\Xi ^{(k)}_{M-1} =  1 + \sum_{j=1}^{M-1} \chi _{j}^{(k,1)}, 
\hspace{0.2cm} M\geq 2 \,,
\label{large_chi} 
\end{equation}
and $\chi _{j}^{(k,1)} = (-1)^{\lambda _1^{(k)} + \lambda _2^{(k)} +
\ldots + \lambda _j^{(k)} }$. 

In the physical frame, we find correspondingly
\begin{eqnarray}
&&\mathbb{E}(\langle \rho_{01}(t_M)\rangle= 
\frac{\rho_{01}(t_0)+\rho_{10}(t_0)}{2\cdot 2^{M-1}} \cdot \\
&&\sum_{k=1}^{2^{M-1}} \left\{ 
\cos \left[ \Xi ^{(k)}_M \omega_0 \Delta t - \delta ^{(k)}(t_M,t_0)
\right] 
e^{-\Gamma^{(k)}(t_M,t_0)} \right\}\: .\nonumber
\end{eqnarray}
Contrary to the result obtained in the absence of noise,
Eq.~(\ref{E_logical}), the additional realization-dependent phase
shift $\delta ^{(k)}(t_M,t_0)$ now remains.  While, on average, this
phase is removed in the limit where $\Delta t \rightarrow 0$, for
finite control rates $\delta ^{(k)}(t_M,t_0)$ may destructively
interfere with the phase gained from the free evolution, potentially
increasing the coherence loss. Identifying specific values of $\Delta
t$ where such harmful interferences may happen for the given RTN
process is not possible, which makes the results for the R-protocol
with finite $\Delta t$ unpredictable in this case.

While the above feature is a clear disadvantage, it is avoided by the
H-protocol. For each realization, the phase accumulated with the free
evolution is completely canceled, so the result in the logical-IP
frame is equal to that in the logical frame: $\mathbb{E}(\langle
\tilde{\rho}_{01}(t_M)\rangle)= \mathbb{E}(\langle
\tilde{\rho}^I_{01}(t_M)\rangle)$.  If access to a classical register
that records the total number of spin flips is also available, this
equivalence between frames may be further extended to the physical
frame. Additionally, as already found in Sect. IIIB, randomized
protocols tend to offer superior stability.

\subsection{Deterministic bursts of switches}

\begin{figure}[thb]
\includegraphics[width=3.4in]{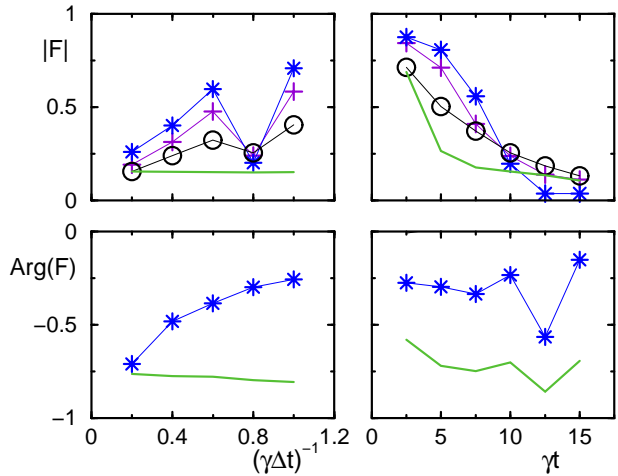}
\caption{(color online) Decoherence rate (upper panels) and phase
offset (lower panels )in the logical-IP frame for a single fluctuator
with $g=1.1$ subjected to a disturbance as given in Eq. (\ref{dist}).
Left panels: Fixed $t_f=10/\gamma $, so $(\gamma \Delta
t)^{-1}=M/10$. Right panels: fixed $\Delta t= 5/(4 \gamma) $.  (Green)
Solid line: Results in the absence of control pulses; (blue) stars:
A-protocol; (black) circles: R-protocol; and (purple) plus:
H-protocol.  The average phase for both R- and H-protocols is zero.
Left panels: Averages are taken over $10^5$ RTN realizations and all
possible $2^M$ pulse realizations.  Right panels: Averages are taken
over $10^4$ RTN realizations and $10^3$ pulse realizations.}
\label{fig:tele_devil}
\end{figure}

Let us illustrate the above statement through an example where the
noisy dynamics of the system is slightly perturbed.  Suppose that,
moving back to the interaction picture, the noise process is now
\begin{equation}
H^I=D(t)\frac{{\rm RTN}(t)}{2}\sigma_z \:,
\label{Ham_tele_devil}
\end{equation}
where 
\begin{equation}
D(t)=(-1)^{\lfloor 4\gamma t/25\rfloor  \lfloor 4\gamma t/5 \rfloor }
\label{dist}\:.
\end{equation}
Physically, $D(t)$ describes a sequence of six instantaneous switches,
equally separated by the interval $5/(4\gamma )$, restarting again at
every instant $25k/(4 \gamma )$, $k$ being an odd number.  This
process may be viewed as bursts of switches of duration $25 /(4\gamma
)$ followed by an interval $25 /(4\gamma )$ of dormancy. The resulting
behavior for $g=1.1$ in the logical-IP frame is depicted in
Fig.~\ref{fig:tele_devil}.

With deterministic control, the rate of noise suppression quickly
improves as the separation between pulses shrinks (left upper panel),
until a certain value, $\Delta t=t_f/8=5/(4\gamma )$, where it
suddenly shows a significant recoil, becoming almost as bad as simply
not acting on the system at all.  Equivalently, by fixing $\Delta
t=5/(4\gamma )$, the performance of the A-protocol becomes very poor
for $t\geq 10/\gamma $ (right upper panel). In practice, detailed
knowledge of the system dynamics might be unavailable, making it
impossible to predict which values of $\Delta t$ might be
adverse. Randomized schemes, on the other hand, are by their own
nature more stable against such interferences.  As seen from the
figure, the R-protocol shows a slower, but also more consistent
improvement as $\Delta t$ decreases, and might therefore be safer in
such conditions.  Notice also that, in terms of coherence preservation
and stability, the H-protocol shows (as intuitively expected) an
intermediate performance between the A- and R-protocols.

{\em To summarize:} In the logical-IP frame, the effects of the RTN
can be reduced not only under deterministic pulses, but also with a
randomized control, though a comparatively shorter pulse separation is
needed in the latter case.  In the logical and physical frames, the
R-protocol, besides showing the poorest performance among the three
considered schemes, may also lead to dangerous interferences between
the qubit frequency and the phase gained from the free evolution. Such
problem, however, does not exist for the H-protocol.  The benefits of
randomization are most clear when limited knowledge about the system
dynamics is available and deterministic control sequences may be
inefficient in avoiding unwanted ``resonances''.  Combining protocols,
where we gain stability from randomness, but also avoid the free phase
evolution, is desirable especially when working in the physical frame.
In this sense, the H-protocol emerges as a promising compromise.

\section{Randomized control of decoherence from a quantum 
bosonic environment}

We now analyze the case of a genuine quantum reservoir, where
decoherence arises from the entanglement between the qubit and the
environment.  The relevant Hamiltonian is given by
Eqs.~(\ref{drift0})-(\ref{drift}) with $\omega_0(t)=\omega_0$ and
$\mu=1$.  In the semiclassical limit, the effects of the interaction
with the bosonic degrees of freedom may be interpreted in terms of an
external noise source whose fluctuations correspond to a Gaussian
random process.

A detailed analysis of deterministic decoherence suppression for this
model was carried out in \cite{Viola1998} (see also~\cite{Dykman79}
for an early treatment of the driven spin-boson model in a nonresonant
monochromatic field and~\cite{Kurizki} for related discussions of
dynamically modified relaxation rates). Here, we discuss how
randomized decoupling performs.

\subsection{Free solution for time-independent interaction 
Hamiltonian}

As in Sect.~IV, we first focus on understanding the controlled dynamics
in a frame that explicitly removes both the control field and the free
evolution due to $H_S\otimes\openone + \openone\otimes H_E$.  Let us
recall some known results related to the uncontrolled
dynamics. We have \cite{Viola1998,Palma}
\begin{equation}
U^I(t,t_0) =\exp \left\{ \frac{\sigma_z}{2} \otimes \sum_k [b^{\dagger}_k
e^{i \omega_k t_0} \xi_k (t-t_0)  -{\rm h.c.}]\right\}\:,
\label{U_IP}
\end{equation}
where
\begin{equation}
\xi_k (\Delta t) = \frac{2g_k}{\omega_k} (1 - e^{i \omega_k \Delta t}) \:.
\label{function}
\end{equation}
Under the standard assumptions that the qubit and the environment are
initially uncorrelated,
\[ \rho_{\rm tot} (t_0) = \rho (t_0)\otimes \rho_E(t_0)\:, \]
and that the environment is in thermal equilibrium at temperature
$T$ [the Boltzmann constant is set =1] 
\[ \rho_E(t_0)= \prod_k  \rho_{E,k}(T) =\prod_k
(1 - e^{\omega_k/T})e^{-\omega_k b_k^{\dagger} b_k/T}\:, \]
the trace over the environment degrees of freedom may be performed
analytically, leading to the following expression for the qubit 
coherence:
\begin{eqnarray}
\rho_{01}^I(t)&&=\rho_{01}(t_0) \prod_k {\rm Tr}_k \{
\rho_{E,k}(T) {\cal D}
[e^{i \omega_k t_0} \xi_k (t-t_0) ]\} \nonumber \\
&&= \rho_{01}(t_0)\exp [-\Gamma (t,t_0)] \:.
\label{decay}
\end{eqnarray}
Here, ${\cal D}(\xi_k) = \exp (b^{\dagger}_k \xi _k - b_k \xi^{*}_k)$
is the harmonic displacement operator of the $k$th bath mode, and the
decoherence function $\Gamma (t,t_0)$ is explicitly given by
\begin{equation}
\Gamma(t,t_0)=\sum_k \frac{|\xi_k (t-t_0)|^2}{2} \coth 
\left(\frac{\omega_k}{2T}\right) \:.
\end{equation}
In the continuum limit, substituting $\sum_k \delta (\omega-\omega_k)|g_k|^2$
by the spectral density $I(\omega)$, one finds
\begin{equation}
\Gamma(t,t_0) =4\int_0^{\infty} d\omega I(\omega) 
\coth \left(\frac{\omega}{2T}\right)
\frac{1-\cos [\omega(t-t_0)]}{\omega^2} \:.
\label{G_nopulse}
\end{equation}
For frequencies less than an ultraviolet cutoff $\omega_c$, 
$I(\omega)$ may be assumed to have a power-law behavior,
\begin{equation}
I(\omega)=\frac{\alpha }{4} \omega^s e^{-\omega/\omega_c} \:.
\label{spectral}
\end{equation}
The parameter $\alpha>0$ quantifies the overall system-bath
interaction strength and $s$ classifies different environment
behaviors: $s=1$ corresponds to the Ohmic case, $s>1$ to the
super-Ohmic and $0<s<1$ to the sub-Ohmic case.

\subsection{Randomly controlled decoherence dynamics: Analytical 
solution and error bound}

Remarkably, the dynamics remains exactly solvable in the presence of
randomized BB kicks.  We focus first on the R-protocol viewed in the
logical-IP frame. Between pulses the evolution is characterized by
Eq.~(\ref{U_IP}), while at $t_j$ Eq.~(\ref{Pclas_IP}) applies. Using
Eq.~(\ref{relation}), the propagator in the logical-IP frame, apart
from an irrelevant overall phase factor, may be finally written as
\begin{equation}
\tilde{U}^I(t_M,t_0)=\exp \hspace*{-.3mm}\left\{\hspace*{-.5mm} 
\frac{\sigma_z}{2} \otimes \hspace*{-.5mm}\sum_k \hspace*{-.6mm}
\left[ b^{\dagger}_k e^{i \omega_k t_0} \eta ^R_k (M,\Delta t)
 - {\rm {\rm h.c.}} \right]\hspace*{-.5mm}
\right\},
\end{equation}
where 
\begin{equation}
\eta ^R_k (M,\Delta t)=\sum _{j=0}^{M-1} \chi_j
e^{i \omega_k j \Delta t} \xi_k (\Delta t) \:.
\label{func}
\end{equation}
Under the uncorrelated initial conditions specified above and thermal
equilibrium conditions, the qubit reduced density matrix is
exactly computed as
\begin{eqnarray}
&\tilde{\rho}_{01}^I(t_M)&=\rho_{01}(t_0) \prod_k {\rm Tr}_k 
\Big\{
\rho_{E,k}(T){\cal D}
[e^{i \omega_k t_0} \eta_k^R(M,\Delta t)]\Big\} \nonumber \\
&&= \rho_{01}(t_0) e^{-\Gamma_{R} (t_M,t_0)}\:.
\label{decoherence}
\end{eqnarray}
Because $\chi_j$ in Eq.~(\ref{func}) can be $\pm 1$ at random,
each element in the sum corresponds to a vector in the complex plane
with a different orientation at every step $\Delta t$.  Thus, the
displacement operator above may be suggestively interpreted as a
random walk in the complex plane.

The decoherence function $\Gamma_{R} (t_M,t_0) $ is now given by
\begin{equation}
\Gamma_{R} (t_M,t_0)=\sum_k \frac{|
\eta_k^R(M,\Delta t)|^2}{2} \coth 
\bigg(\frac{\omega_k}{2T}\bigg)\:,
\end{equation}
which, in the continuum limit, becomes
\begin{widetext}
\begin{eqnarray}
\Gamma_{R} (t_M,t_0)=4\int_0^{\infty} 
d\omega I(\omega) \coth \left( \frac{\omega}{2T}\right)
\frac{1-\cos (\omega \Delta t)}{\omega^2}
\bigg[ M + 2\sum_{j=1}^{M-1} \cos (j \,\omega \Delta t)\sum _{l=0}^{M-j-1}
\chi _l \, \chi _{l+j}  \bigg] \:.
\label{main_result}
\end{eqnarray}
\end{widetext}

The decoherence behavior under the A-protocol is obtained by letting
$\chi_j=(-1)^j$. We then recover the result of deterministically
controlled decoherence~\cite{Viola1998}, which may be further
simplified as~\cite{Shiokawa2002},~\cite{Gheorghiu},
\begin{widetext}
\begin{eqnarray}
\Gamma_{D}(t_M,t_0)=4\int_0^{\infty} 
d\omega I(\omega ) \coth 
\left( \frac{\omega}{2T} \right)
\frac{1 - \cos [\omega (t_M-t_0)]}{\omega^2} 
\tan ^2 \left( \frac{\omega \Delta t}{2} \right)\:.
\label{determ}
\end{eqnarray}
\end{widetext}

Before proceeding with a numerical comparison between
Eqs.~(\ref{main_result})-(\ref{determ}), some insight may be gained
from an analytical lower bound for the average $\mathbb{E}(\exp
[-\Gamma_{R} (t_M,t_0)])$.  According to Jensen's inequality,
$\mathbb{E} (f(x)) \geq f(\mathbb{E}(x))$ for any convex function
$f$. Using this and the fact that
\[ \mathbb{E}(\chi _l \chi _{l+j}) = \mathbb{E} 
\left((-1)^{\lambda _{l+1} +\lambda _{l+2}+\ldots + \lambda _{l+j}}
\right)=0\:, \]
we have the following lower bound 
\begin{widetext}
\begin{eqnarray}
\label{lower_bound}
\hspace*{5mm}\mathbb{E}\big( \exp [-\Gamma_{R} (t_M,t_0)] \big) \geq
\exp [- \mathbb{E} \big( \Gamma_{R} (t_M,t_0) \big)] \:,\hspace*{10mm} 
\nonumber \\
\mathbb{E} \big( \Gamma_{R} (t_M,t_0) \big)  = 
 4 M \int_0^{\infty} 
d\omega I(\omega) \coth \left( \frac{\omega}{2T}\right)
\frac{1-\cos (\omega \Delta t)}{\omega^2}  \:. 
\end{eqnarray}
\end{widetext}

\begin{figure}[b]
\includegraphics[width=3.45in]{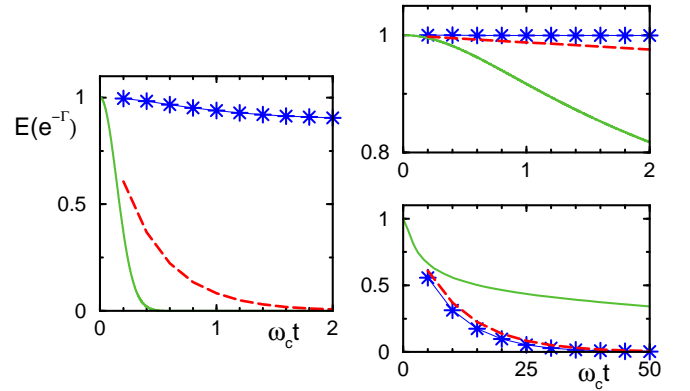}
\caption{(color online) Decoherence rate from a bosonic Ohmic bath.
Here and in the following figures, time is measured in units of
$T^{-1}$, $\alpha=0.25 $, $\omega_c=100$.  Left panel: $T=10^2
\omega_c$ and $\omega_c \Delta t=0.1$. Right panels:
$T=10^{-2}\omega_c$, top: $\omega_c \Delta t=0.1$ and bottom:
$\omega_c \Delta t=2.5$.  (Green) Solid line: No control; (blue)
stars: A-protocol; (red) dashed line: Lower bound,
Eq.~(\ref{lower_bound}). }
\label{fig:bound}
\end{figure}

In Fig.~\ref{fig:bound} we compare the coherence decay corresponding
to the absence of control (\ref{G_nopulse}), to the A-protocol
(\ref{determ}) and to the lower bound (\ref{lower_bound}).  Two
limiting cases of high and low temperature, $T\gg \omega_c$ and $T\ll
\omega_c$, are considered.  The high temperature limit corresponds to
an effectively classical bath, where the properties of the environment
are dominated by thermal fluctuations.  In the absence of control,
decoherence is very fast on the time scale determined by the bath
correlation time $\tau_c=\omega_c^{-1}$, hence coherence preservation
requires very short intervals between pulses.  The A-protocol shows
the best performance. The actual randomized performance may, however,
be significantly better than the lower bound in this temperature
regime, though they never surpass the deterministic case [see next
subsection].

In the case of low temperature, or fully quantum bath, decoherence is
much slower and a richer interplay between thermal and vacuum
fluctuations occurs.  Larger values of $\Delta t$ may then be analyzed
before total coherence loss takes place.  The interesting phenomenon
of decoherence acceleration~\cite{Viola1998,Tombesi99}, which may
happen when $\omega_c \Delta t>1$, may now be observed.   
For short $\Delta t$, the A-protocol is again more efficient, though
not significantly better than the lower bound.  For large $\Delta t$,
pulses induce destructive interference and the A-protocol performs
even worse than the lower bound. In such situation the best option is
simply not to act on the system.

When $\omega_c \Delta t\lesssim 1$, some general insight may be gained
by comparing appropriate limits of the lower bound and the
deterministic decoherence function. First, by Taylor-expanding up to
second-order in $\Delta t$ we have
\begin{eqnarray}
\label{new_lower_bound}
&&\mathbb{E} \big( \Gamma_{R} (t_M,t_0) \big) \approx \\
&&  2 (t_M - t_0) \Delta t \int_0^{\infty} 
d\omega \, I(\omega) \coth \left( \frac{\omega}{2T}\right)
\nonumber \:, 
\end{eqnarray}
whereas 
\begin{eqnarray}
\label{new_determ}
&&\Gamma_{D}(t_M,t_0)\approx \\
&&\Delta t^2\int_0^{\infty} 
d\omega I(\omega ) \coth 
\left( \frac{\omega}{2T} \right)
\{1 - \cos [\omega (t_M-t_0)] \}  \nonumber \:.
\end{eqnarray}
Therefore, in the limit of very short $\Delta t$, the lower bound
approaches the ideal situation of total suppression of decoherence
linearly in $\Delta t$, while for the A-protocol this occurs
quadratically.

This analysis may be further extended by studying the two limits of
Eq.~(\ref{new_lower_bound}) with respect to temperature.  Considering
the spectral density of Eq.~(\ref{spectral}), we have
\begin{eqnarray*}
&&  T\gg \omega_c: \hspace{0.5cm} 
\mathbb{E}\big( \Gamma_{R} (t_M,t_0) \big) = {\cal O}\left(
\alpha T \omega_c^s (t_M - t_0) \Delta t \right) \:,\\
&& T\ll \omega_c: \hspace{0.5cm}
\mathbb{E}\big( \Gamma_{R} (t_M,t_0) \big) = {\cal O}\left(
\alpha \omega_c^{s+1} (t_M - t_0) \Delta t \right) \:.
\end{eqnarray*}
Thus, a sufficient condition under which random control avoids
decoherence is 
\begin{eqnarray*}
&& T\gg \omega_c: \hspace{0.5cm} 
\alpha T\omega_c^s (t_M-t_0) \Delta t \ll 1 \:,\\
&& T\ll \omega_c: \hspace{0.5cm} 
\alpha \omega_c^{s+1} (t_M-t_0) \Delta t \ll 1\:. 
\nonumber
\end{eqnarray*}
This should be compared with the general bound given in Theorem 2
of~\cite{Viola2005Random}.  We will see in Sect.~V.D that, in the
physical frame, the qubit frequency $\omega_0$ also plays an important
role.

A similar analysis for the A-protocol may be effected using
Eq.~(\ref{new_determ}).  For $\omega_c (t_M-t_0)\ll 1$, the
decoherence function decays quadratically in time, giving
\begin{eqnarray}
&& T\gg \omega_c: \hspace{0.5cm} 
\Gamma _D(t_M,t_0)  = {\cal O}\left(
\alpha T\omega_c^{s+2} (t_M-t_0)^2 \Delta t^2 \right)\:, 
\nonumber \\
&& T\ll \omega_c: \hspace{0.5cm} 
\Gamma _D(t_M,t_0)  = {\cal O}\left( 
\alpha \omega_c^{s+3} (t_M-t_0)^2 \Delta t^2\right)
\:, \nonumber
\end{eqnarray}
while for $\omega_c (t_M-t_0)\gg 1$ the decoherence function becomes
independent of $t_M-t_0$, and we get
\begin{eqnarray}
&& T\gg \omega_c: \hspace{0.5cm} 
\Gamma _D(t_M,t_0)  = {\cal O}\left( 
\alpha T\omega_c^s \Delta t^2 \right)\:,
\nonumber \\
&& T\ll \omega_c: \hspace{0.5cm} 
\Gamma _D(t_M,t_0)  = {\cal O}\left(
\alpha \omega_c^{s+1} \Delta t^2\right)
\:. \nonumber
\end{eqnarray}
These should be compared with the error bound of Theorem 3
in~\cite{Viola2005Random}.  Based on these considerations, random
pulses may hardly be expected to outperform deterministic controls in
the limits discussed above. Still, it remains interesting to
quantitatively see what the actual performance is for intermediate
$\Delta t$ and/or hybrid schemes, for instance with respect to
acceleration.  Moreover, further changes may be expected when some
time dependence exists in the system parameters, for instance in the
coupling strength to the environment [see Sect.~V.E].

\subsection{Randomly controlled decoherence
dynamics: Numerical results}

Based on the exact result of Eq.~(\ref{main_result}), we now present a
comparison of the average decoherence suppression achievable by the
protocols described in Sect.~II.  A fixed time $t_f$ divided into an
increasing number $M$ of intervals $\Delta t$ is considered.

\begin{figure}[htb]
\includegraphics[width=2.8in]{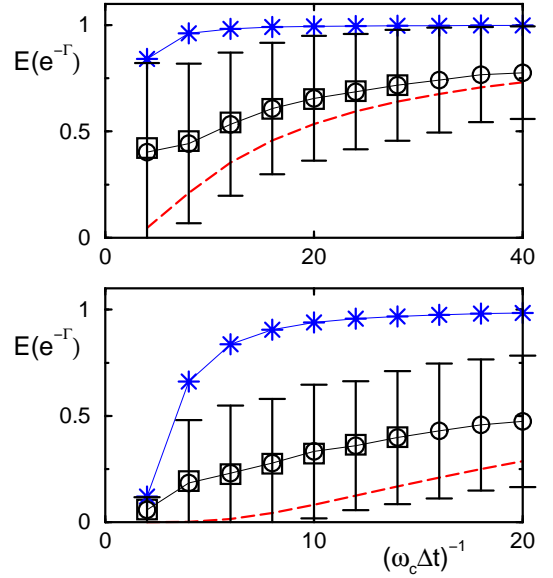}
\caption{(color online) Decoherence rate for a high temperature Ohmic
bath, $T=10^2\omega_c$.  Upper panel: $\omega_c t_f=0.5$; lower panel:
$\omega_c t_f=1$.  (Blue) stars: A-protocol; (black) circles: Average
over $10^3$ realizations and respective standard deviations for the
R-protocol~\cite{standard}; (black) squares: Expectation value (taken
over all $2^M$ realizations) for the R-protocol; (red) dashed line:
Lower bound. }
\label{fig:high}
\end{figure}

Fig.~\ref{fig:high} compares the average $\mathbb{E}(\exp [-\Gamma_{R}
(t_M,t_0)])$ in the limit of high temperature, $T=10^2\omega_c$, for a
system evolving under the A- and R-protocols.  For the fixed times
chosen, $\omega_c t_f=0.5$ (upper panel) and $\omega_c t_f=1$ (lower
panel), the coherence element has already practically disappeared and
cannot be seen in the figure, while the A-protocol is able to recover
it even for very few cycles. The values of $\exp[-\Gamma_{R}
(t_M,t_0)]$ for different realizations are widely spread between the
worst case corresponding to all $\lambda$'s=0 and the efficient
realizations involving several spin flips. As a consequence, the
average converges to 1 slowly and has a large standard deviation
\cite{standard}.  Notice, however, that it is significantly better
than the lower bound.

\begin{figure}[htb]
\includegraphics[width=2.8in]{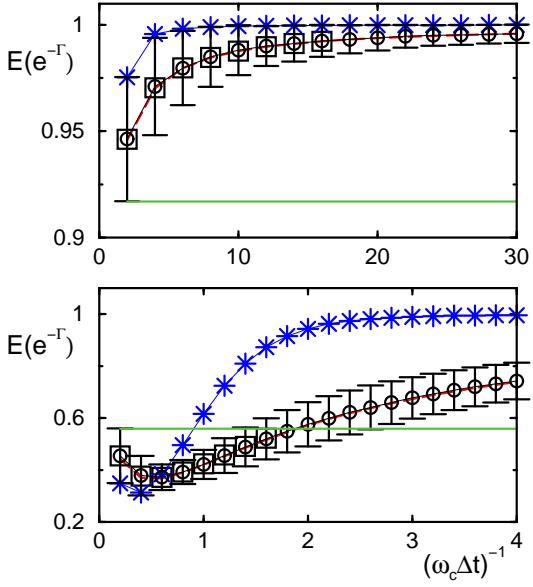}
\caption{(color online) Decoherence rate for a low temperature Ohmic
bath, $T=10^{-2}\omega_c$.  Upper panel: $\omega_c t_f=1$; lower
panel: $\omega_c t_f=10$. (Green) Solid line: No control; (blue)
stars: A-protocol; (black) circles: Average over $10^3$ realizations
and respective standard deviations for the R-protocol; (black)
squares: Expectation value for the R-protocol.}
\label{fig:low}
\end{figure}

The results in the low-temperature limit, $T=10^{-2}\omega_c$, are
shown in Fig.~\ref{fig:low}. Here, thanks to the fact that decoherence
is overall slower, longer evolution times may be chosen: $\omega_c
t_f=1$ (upper panel) and $\omega_c t_f=10$ (lower panel). For the
latter choice, in particular, when $M \lesssim 10$, decoherence
enhancement occurs and, interestingly, the results for the A-protocol
are {\em worse} than those for the R-protocol.  However, it takes a
much smaller $\Delta t$ for the R-pulses to finally cross the line
that separates enhancement from decoherence reduction.  Notice also
that the values of $\exp [-\Gamma_{R} (t_M,t_0)]$ for different
realizations are not so spread and the standard deviations are
narrower than in the high temperature limit.  In addition, the average
over realizations is very close to the lower bound, to the point that
they cannot be distinguished in the figure.

\begin{figure}[htb]
\includegraphics[width=2.8in]{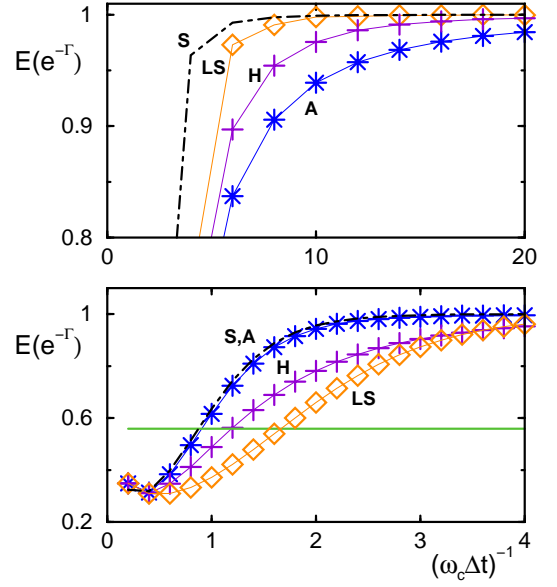}
\caption{(color online) Decoherence rate. Upper panel: High
temperature Ohmic bath, $T=10^{2}\omega_c$, and $\omega_c
t_f=1$. Lower panel: Low temperature Ohmic bath, $T=10^{-2}\omega_c$,
and fixed time $\omega_c t_f=10$.  (Green) Solid line: No control;
(blue) stars: A-protocol; (purple) plus: H-protocol; (orange)
diamonds: LS-scheme; and (black) dot-dashed line: S-protocol.  Average
for the H-protocol is taken over $10^3$ realizations.  }
\label{fig:sym}
\end{figure}

We now extend our comparison to the three remaining protocols of
Fig.~\ref{fig:scheme}, see Fig.~\ref{fig:sym}.  We choose a high
temperature bath with $\omega_c t_f=1$ (upper panel) [low temperature,
in this case, leads to similar results], and a low temperature bath
with $\omega_c t_f=10$ (lower panel).  The S-protocol shows the best
performance, which is evident in the upper panel, but hardly
perceptible in the lower one.  Due to the different rearrangement of
the time interval between pulses for this protocol, it does not
correspond to any of the $2^M$ realizations of random pulses as
considered here and represent a special scheme separated from the
others.  The performance of the LS-protocol, which has half the number
of $\pi$-pulses used in the A-protocol, turns out to be better in all
cases of a high temperature bath, but worse for a fully quantum bath
with large $\omega_c t_f$.  This explains why the H-protocol, which
combines symmetrization and randomness, also performs better than the
A- in the high temperature limit.

{\em To summarize:} In terms of performance, we have in decreasing order: S-,
LS-, H-, A-, R-protocols for high temperature;
and S-, A-, H-, LS-, R-protocols for low temperature 
once the number of pulses are sufficient to start slowing down
decoherence.  Different protocols become again, as expected,
essentially equivalent in the limit $\Delta t\rightarrow 0$.  For
finite pulse separations, in the considered case of a time-independent
Hamiltonian, it is always possible to identify a deterministic
protocol showing the best performance.  However, if a balance is
sought between good performance and protocols minimizing the
required number of pulses, then the H-protocol again emerges as an
interesting compromise.  Note, in particular, that the latter
outperforms the standard A-protocol in some parameter regimes.

\subsection{Randomized decoupling in the physical frame}

We now investigate under which conditions decoherence suppression is
attainable in the physical frame, when the system is subjected to
randomized control.  Because, in this frame, the qubit natural
frequency plays an important role, random decoupling also depends on
how small $\Delta t$ can be made with respect to
$\tau_0=\omega_0^{-1}$.

The reduced density matrix is obtained following the same steps
described so far, but in order to retain the effects of the system
Hamiltonian, the transformation into the interaction picture is now
done with respect to the environment Hamiltonian only -- hence the
superscript $IE$. Upon tracing over the environment degrees of freedom,
we are left with the reduced density operator in the Schr\"odinger
picture.

The unitary operator between pulses is 
\begin{eqnarray}
&&U^{IE}(t_{j+1}, t_j)= \exp \left(-i \frac{\omega_0 }{2}
\sigma_z \Delta t\right)  \\
&& \hspace{0.5cm}
\exp \bigg\{ \frac{\sigma_z }{2} \otimes
\sum _k
\left[
\xi_k (\Delta t) b^{\dagger}_k e^{i\omega_k t_j} - {\rm {\rm h.c.}}
\right] \bigg\}\:,
\label{U_IPbath} \nonumber
\end{eqnarray}
while at $t_j$ it is given by
\begin{equation}
P_j^{IE}=\exp \left[-i\lambda _j \frac{\pi}{2} \sigma _x \right]\:. 
\label{P_IPbath}
\end{equation}
By additionally moving to the logical frame we get
\begin{eqnarray}
&&\tilde{U}^{IE} (t_M,t_0) = 
\exp \bigg[ -i \frac{\omega_0}{2} \sigma_z \Delta t
\sum_{j=0}^{M-1} \chi _j \bigg]  \\
&& \hspace{0.5cm}
\exp \bigg\{ \frac{\sigma_z}{2} \otimes \sum_k
 \left[ b^{\dagger}_k e^{i \omega_k t_0} \eta ^R_k (M,\Delta t)
 - {\rm {\rm h.c.}} \right]
\bigg\} \:. \nonumber
\end{eqnarray}
Tracing over the environment and taking the expectation over control 
realizations leads to the coherence element in the logical frame:
\begin{equation}
\mathbb{E}(\tilde{\rho}_{01}(t_M))=
\frac{\rho_{01}(t_0)}{2^{M-1}} \sum_{k=1}^{2^{M-1}} 
\cos [\Xi _{M-1}^{(k)} \omega_0 \Delta t]
e^{-\Gamma_{R}^{(k)} (t_M,t_0)}\:,
\end{equation}
where $\Xi _{M-1}^{(k)}$ is given by Eq.~(\ref{large_chi}).  Thus, in
addition to the decoherence described as before by
Eq.~(\ref{main_result}), we now have ensemble dephasing due to the
fact that each realization carries a different phase factor
proportional to $\omega_0$.

\begin{figure}[b]
\includegraphics[width=2.7in,height=3.1in]{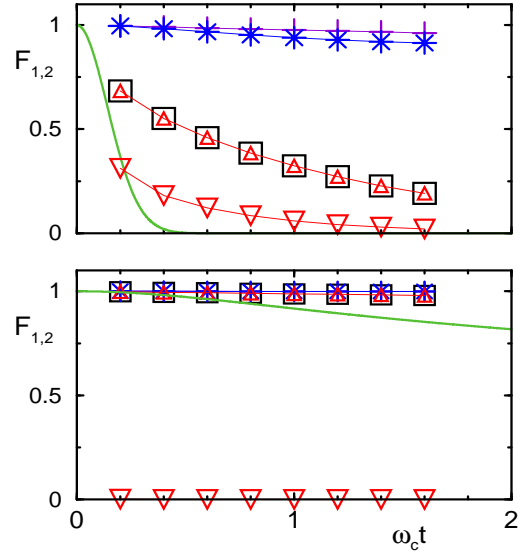}
\caption{(color online) Ratios ${\rm F}_1(t)$ and ${\rm F}_2(t)$ in
the logical and logical-IP frames, respectively.  A fixed time
interval $\Delta t=1/(10 \omega_c)$ is taken.  Upper panel: High
temperature Ohmic bath, $T=10^{2}\omega_c$, lower panel: Low
temperature Ohmic bath, $T=10^{-2}\omega_c$. (Green) Solid line: No
control; (blue) stars: A-protocol; (purple) plus: H-protocol; (black)
squares: R-protocol in the logical-IP frame; (red) up triangles:
R-protocol in the logical frame with small frequency $\omega _0 \Delta
t = 10^{-3}$; (red) down triangles: R-protocol in the logical frame
with large frequency $\omega _0 \Delta t = \pi/2$.  Average performed
over all realizations.}
\label{fig:IPbath}
\end{figure}

The results for the ratios 
\begin{equation}
{\rm F}_1(t_M) = \frac{\mathbb{E}(\tilde{\rho}_{01}(t_M))}{\rho_{01}t_0)}
\hspace{0.3cm} {\rm and} \hspace{0.3cm}
{\rm F}_2(t_M) = \frac{\mathbb{E}(\tilde{\rho}^I_{01}(t_M))}{\rho_{01}(t_0)}
\end{equation}
in the logical and logical-IP frames for the system, respectively, are
summarized in Fig.~\ref{fig:IPbath}, where $\Delta t$ is fixed and the
system is observed  at different times. Both a  high temperature and a
low  temperature   scenario  are  considered.   The   phase  for  each
realization in  the logical frame is mostly  irrelevant when $\omega_0
\ll 1/\Delta t$. The outcomes  of the average over all realizations in
both   frames  are   then  comparable,   independently  of   the  bath
temperature.   The situation changes  dramatically when  the spin-flip
energy becomes  large, the  worst scenario corresponding  to $\omega_0
=k\pi/(2 \Delta t)$, with  $k$ odd. Here, because $\Xi _{M-1}^{(k)}$ is
an  even number,  each  realization  makes a  positive  or a  negative
contribution to the average, which may therefore be very much reduced.
Such  destructive quantum  interference is  strongly dependent on the
bath temperature.

Among all random pulse realizations, the most effective at suppressing
decoherence are those belonging to the smaller ensemble of the
H-protocol.  None of them carries a phase, so they always make large
positive contributions to the total ensemble average. In a high
temperature bath, the realizations that can make negative
contributions have often tiny values of $e^{-\Gamma_{R}(t_M,t_0)}$,
which explains why even in the extreme case of $\omega_0
=k\pi/(2\Delta t)$ the R-protocol can still lead to some decoherence
reduction. In a low temperature bath, on the other hand, decoherence
is slower and for the time considered here, the values of
$e^{-\Gamma_{R}(t_M,t_0)}$ for all realizations are very close, which
justifies their cancellation when $\omega_0 =k\pi/(2\Delta t)$.
 
In the physical frame, the average for the density matrix depends on
the initial state of the system as
\begin{eqnarray*}
&&\mathbb{E}(\rho_{01}(t_M))=
\frac{\rho_{01}(t_0) + \rho_{10}(t_0)}{2\cdot 2^{M-1}}  \nonumber \\
&&
\hspace*{2.1cm} 
\sum_{k=1}^{2^{M-1}} 
\cos [\Xi _{M}^{(k)} \omega_0 \Delta t]
e^{-\Gamma_{R}^{(k)} (t_M,t_0)}\:.
\end{eqnarray*}
As already discussed in Sect.~III, the problem associated with 
population inversion may be avoided if a classical register is used 
to record the actual number of spin flips.

{\em To summarize:} Two conditions need to be satisfied for the
R-protocol to become efficient in reducing decoherence: $\omega_c
\Delta t\ll 1$ and also $\omega_0 \Delta t\ll 1$.  Notice, however,
that when randomness and determinism are combined in a more elaborated
protocol, such as the H-protocol, no destructive interference due to
$\omega_0$ occurs. In addition, the hybrid scheme is still capable to
outperform the A-protocol in appropriate regimes.

\subsection{Time-dependent coupling Hamiltonian}

As a final example, imagine that the coupling parameters $g_k(t)$
between the system and the environment are time dependent and let us
for simplicity work again in the logical-IP frame.  The total
Hamiltonian is given by Eqs.~(\ref{drift0})-(\ref{drift}) with
$\omega_0(t)=\omega_0$ and $\mu=1$.  Two illustrative situations are
considered: $g_k(t)$ changes sign after certain time intervals, or it
periodically oscillates in time.

\subsubsection{Instantaneous sign changes}

\begin{figure}[tb]
\includegraphics[width=2.7in]{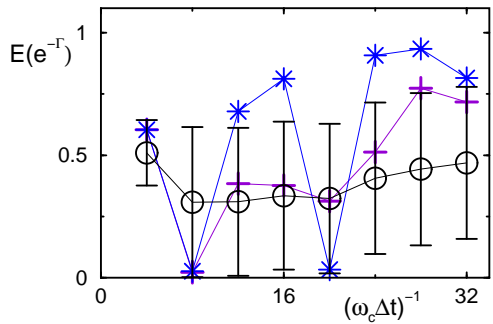}
\caption{(color online) Decoherence rate for a high temperature Ohmic
reservoir, $T=10^{2}\omega_c$, with alternating couplings.  (Blue)
stars: A-protocol; (black) circles: R-protocol; (purple) plus:
H-protocol.  The interval considered is $t_f=1/\omega_c$.  Averages
taken over all possible realizations. }
\label{fig:DEVIL_quantum}
\end{figure}

Suppose that $g_k(t)=g_k D(t)$, where
\begin{equation}
D(t)=(-1)^{\lfloor \frac{10 \omega _c t}{3} \rfloor }
\label{devil_bath1}
\end{equation}
describes instantaneous sign changes of the coupling parameter after
every interval $3/(10 \omega_c)$.  For a high temperature bath and a
fixed time $t_f=1/\omega_c$, Fig.~\ref{fig:DEVIL_quantum} shows that
the results for the A-protocol exhibit a drastic drop when $\Delta
t=t_f/4$ and $\Delta t=t_f/10$.  This is due to the fact that some of
sign changes happen very close to or coincide with some of
$\pi$-pulses of the deterministic sequence, canceling their effect.
In contrast, the occurrence of spin flips in randomized schemes is
irregular, so that the latter are more protected against such
``resonances'' and steadily recover coherence as $\Delta t$ decreases,
even though at a slower pace.
Note that when dealing with the S- or LS-protocols, the same sort of
recoil should be expected for different time dependences and different
values of $\Delta t$.

\subsubsection{Periodic modulation}

\begin{figure}[htb]
\includegraphics[width=3.1in]{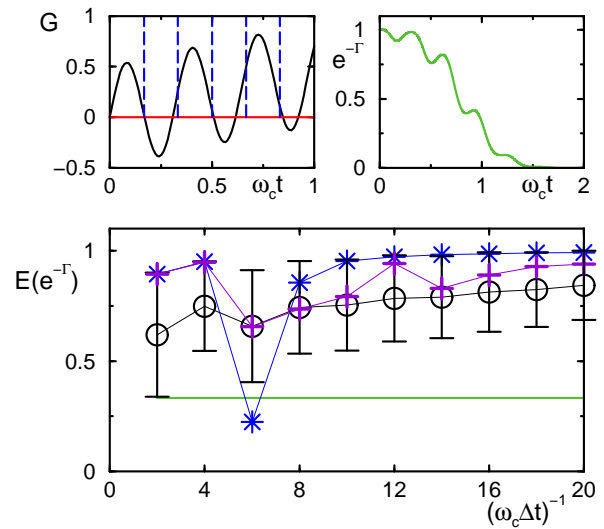}
\caption{(color online) Upper left panel: Function $G(t)= \cos(p \,\pi
\omega_c t) \sin(q\, \pi \omega_c t)$, for $p=2.95 $ and $q=3.25 $.
Upper right panel: decoherence rate in the absence of control.  Lower
panel: Decoherence rate for a fixed time interval $\omega_c t_f=1$. A
high temperature Ohmic bath, $T=10^2 \omega_c$, is considered.
(Green) Solid line: No control; (blue) stars: A-protocol; (black)
circles: R-protocol; (purple) plus: H-protocol. Averages taken over
$10^3$ realizations.  Standard deviations for the R-protocol are
shown.  }
\label{fig:time_bath}
\end{figure}

Assume that the coupling parameter is given by $g_k(t)=g_kG(t)$, where
$G(t)=\cos(p \,\pi\omega_c t) \sin(q \,\pi\omega_c t)$ and $|p-q|$ is
small.  This function has two superposed periodic behaviors, one with
a long period and the other fast oscillating. The fast oscillations
are shown in the left upper panel of Fig.~\ref{fig:time_bath}.

We consider a high temperature bath, $T=10^2 \omega_c$.  The right
upper panel shows the qubit decoherence in the absence of pulses. The
oscillations in the decay rate are related to the oscillations in the
interaction strength between the system and the bath. In the lower
panel, we fix a time $t_f=1/\omega_c $ and compare the decoherence
rate for the cases of absence of control, A-, H- and R-protocols.  When
$\Delta t=t_f/6$ the result for the A-protocol suddenly becomes even
worse than not acting on the system.  Random pulses, on the contrary,
do not show any significant recoil.  The reason for the inefficiency
of the A-protocol when $M=6$ becomes evident from the left upper panel
of Fig.~\ref{fig:time_bath}.  Vertical dashed lines indicate where the
pulses occur.  They mostly coincide with the instants where $G(t)$
also changes sign.  For the LS-protocol, similar unfavorable
circumstances happen for different values of $\Delta t$ and similar
behaviors should be expected for other deterministic protocols and
functions $g_k(t)$.

{\em To summarize:} The above examples again reinforce the idea of
enhanced stability of randomized controls, and suggest that
randomization might represent a safer alternative in reducing
decoherence when limited knowledge about the system-bath interaction
is available.

\section{Conclusion}

\subsection{Summary}

A quantitative comparison between deterministic and randomized control
for the most elementary target system, consisting of a single
(isolated or open) qubit, was developed in different frames.  The main
conclusions emerging from this study may be summarized as follows.

First, it is always possible to identify conditions under which purely
random or hybrid schemes succeed at achieving the desired level of
dynamical control.  Frame considerations play an important role in
specifying such conditions, satisfactory performance in a given frame
being ultimately determined by a hierarchy of time scales associated
with all the dynamical components in the relevant Hamiltonian. While
all protocols become essentially equivalent in the limit of
arbitrarily fast control, the behavior for finite pulse separation is
rich and rather sensitive to the details of the underlying dynamics.
As a drawback of pure random design, ensemble average tends to
introduce, in general, additional phase damping, which may be however
circumvented by combining determinism and randomness within a hybrid
design.

Second, for time-independent control settings in this simple system,
it was always possible to identify a deterministic protocol with best
performance.  While deterministic schemes ensuring accurate averaging
of a known interaction always exist in principle~\cite{HaeberlenBook},
such a conclusion remains to be verified under more general
circumstances, in particular access to a {\em restricted} set of
control operations.  The hybrid protocol proved superior to the pure
random schemes, as well as to standard asymmetric schemes in certain
situations.

Third, for time-varying systems, randomized protocols typically allow
for enhanced stability against parameter variations, which may
severely hamper the performance of deterministic schemes.  Pure random
design tends to perform better than hybrid in this respect, both
choices, however, improving over purely cyclic controls under
appropriate conditions.

Overall, hybrid design emerges as a preferred strategy for merging
advantageous features from different protocols, thereby allowing to
better compromise between conflicting needs.

\subsection{Outlook}

From a conceptual standpoint, it is intriguing to realize that
complete suppression of decoherence remains possible, in principle, by
purposefully introducing a probabilistic component in the underlying
control, and perhaps surprising to identify cases where this leads to
improved efficiency over pure deterministic methods.

In a broader context, however, it is worth mentioning that the
philosophy of recognizing a beneficial role of randomness in physical
processes has a long history.  Within NMR, the stochastic averaging of
intermolecular interactions in gases and isotropic liquids due to
random translational and re-orientational motions may be thought of as
a naturally occurring random self-decoupling
process~\cite{HaeberlenBook}.  In spectroscopic applications of
so-called stochastic NMR and stochastic magnetic-resonance
imaging~\cite{StochNMR}, spin excitation via trains of weak RF pulses
randomly modulated in amplitude, phase, and/or frequency are used to
enhance decoupling efficiencies over a broader frequency bandwidth
than attainable otherwise.  Even more generally, the phenomenon of
stochastic resonance~\cite{Gammaitoni,Viola2000StochRes} is
paradigmatic in terms of pointing to a constructive role of noise in
the transmission of physical signals.  Within QIP, strategies aimed at
taking advantage of noise and/or stochasticity have been considered in
contexts ranging from quantum games~\cite{Johnson}, to quantum
walks~\cite{Kendon}, dissipation-assisted quantum
computation~\cite{Beige}, as well as specific
coherent-control~\cite{Mancini2002} and quantum
simulation~\cite{Bremner2004} scenarios.  Yet another suggestive
example is offered by the work of Prosen and \v{Z}nidari\v{c}, who
have shown how static perturbations characterizing faulty gates may
enhance the stability of quantum algorithms~\cite{Prosen}.  More
recently, as mentioned, both pure random~\cite{Kern2004} and
hybrid~\cite{Kern2005} active compensation schemes for static coherent
errors have been proposed.  While it is important to stress that none
of the above applications stem from a general {\em control-theoretic
framework} as developed in~\cite{Viola2005Random}, it is still
rewarding to fit such different examples within a unifying
perspective.

Our present analysis should be regarded as a first step toward a
better understanding and exploitation of the possibilities afforded by
randomization for coherent and decoherent error control.  As such, it
should be expanded in several directions, including more realistic
control systems and settings, and fault-tolerance considerations.
While we plan to report on that elsewhere, it is our hope that our
work will stimulate fresh perspectives on further probing the
interplay between the field of coherent quantum control and the world
of randomness.

\begin{acknowledgments}
L. V. warmly thanks Manny Knill for discussions and feedback during
the early stages of this project. The authors are indebted to an
anonymous referee for valuable suggestions. L. F. S. gratefully
acknowledges support from Constance and Walter Burke through their
Special Projects Fund in Quantum Information Science.
\end{acknowledgments}

\appendix
\section{Control Hamiltonian}

The control Hamiltonian is designed according to the intended
modification of the target dynamics in a desired frame. Throughout
this work, our goal has been to freeze the system evolution by
removing any phase accumulated due to the unitary evolution, as well
as avoiding nonunitary ensemble dephasing and decoherence.  As
clarified below, this requires the use of {\em identical $\pi$-pulses
in the physical frame}. This condition may be relaxed at the expense
of no longer refocusing the unitary evolution.

Let the control of the system be achieved via the application of an
external alternating field (e.g., a radiofrequency magnetic field),
\begin{equation}
H_c(t)=\sum_{j} H_c^{(j)}(t)=\sum_{j} V^{(j)} (t)
\cos [\omega t + \varphi_j (t)] \sigma_x \:,
\label{Hc_cos}
\end{equation}
where the carrier is tuned on resonance with the qubit central
frequency, $\omega = \omega_0$.  As described in the text,
$V^{(j)}(t)=V[\Theta (t-t_j) - \Theta (t-t_j-\tau)]$ and each pulse
happens at $t_j$, having duration $\tau $ and amplitude $V$. Upon
invoking the rotating wave approximation (RWA), hence neglecting the
counter-rotating terms $\sigma_- \exp [-i(\omega_0 t + \varphi_j
(t))]$ and $\sigma_+ \exp [i(\omega_0 t + \varphi_j (t))]$ in
(\ref{Hc_cos}), the control Hamiltonian given in the main text is
found.  The function $\varphi_j (t)$ characterizes the phase
properties of the pulses we deal with. We compare two relevant
possibilities:

\vskip 0.1 cm
\begin{flushleft}
{\rm (i) $\varphi_j (t)=-\omega_0 t_j$ for each $j$: This means that
the pulses are identical in the physical frame, as used in this work.}

\vskip 0.1 cm {\rm (ii) $\varphi_j (t)=0$ for all $j$: This means that
the pulses are identical in the physical frame only if separated in
time by a multiple of $2\pi$.}
\end{flushleft}

The propagator corresponding to the above choices may be in general
obtained by seeking a transformation which removes the time dependence
of $H^{(j)}_c(t)$ within each pulse.  A transformation to an {\em
absolute} frame rotating with the carrier frequency, which on
resonance is identical with the interaction picture, leads to
\begin{eqnarray}
H_c^{I\, (j)} (t)&=&\exp \left[i \omega_0 t\sigma_z /2 \right] 
H_c^{(j)}(t)
\exp \left[-i \omega_0 t\sigma_z /2 \right]
\nonumber \\
&=& V^{(j)}(t) \exp 
\left[-i \frac{\varphi_j(t)}{2}\sigma_z\right]
\sigma_x \exp \left[i \frac{\varphi_j(t)}{2}\sigma_z \right]\:. \nonumber
\end{eqnarray}
Thus, the choice $\varphi_j (t)=0$ corresponds to pulses which are
translationally invariant in time in this frame.  In case (i), the
above transformation does not remove time dependence, which would
instead be accomplished by moving to a {\em relative} rotating frame
via a rotation $U_z(t-t_j)=\exp[i\omega_0(t-t_j)\sigma_z/2]$.  
From the
above expression, the interaction-picture propagators for an
instantaneous $\pi$-pulse applied at $t_j$ are found, respectively,
as 
\begin{eqnarray}
{\rm (i)} &\:& P_j^I = \exp \left[i \frac{\omega_0 t_j}{2} \sigma_z \right]
\exp \left[-i \frac{\pi}{2} \sigma_x  \right] 
\exp \left[-i \frac{\omega_0 t_j}{2} \sigma_z \right], \nonumber \\
{\rm (ii)} &\:& P_j^I = \exp \left[-i \frac{\pi}{2} \sigma_x \right].
\nonumber 
\end{eqnarray}
We can then return to the Schr\"odinger picture using the relation
\[ P_j=\exp [-i \omega_0 t_j \sigma_z /2] P_j^I  
\exp [i \omega_0 t_j \sigma_z /2]\:, \]
leading to the propagators
\begin{eqnarray}
{\rm (i)} &\:& P_j = \exp \left[-i \frac{\pi}{2} \sigma_x \right],
\label{P_varphi} \\
{\rm (ii)} &\:& P_j = \exp \left[-i \frac{\omega_0 t_j}{2} \sigma_z  \right]
\exp \left[-i \frac{\pi}{2} \sigma_x  \right] 
\exp \left[i \frac{\omega_0 t_j}{2} \sigma_z \right]\:. \nonumber 
\end{eqnarray}
Thus, pulses with $\varphi_j (t)=-\omega_0 t_j$ are confirmed to be
translationally invariant in time in the physical frame, as directly
clear from the dependence $\omega_0 (t-t_j)$ in (\ref{Hc_cos}).

The difference between the two choices to the control purposes 
becomes evident by considering the A-protocol on the isolated qubit. 
From (\ref{P_varphi}),
the propagators $U(t_2,t_0)=P_2 U_0(t_2,t_1) P_1 U_0(t_1,t_0)$ in the
physical frame are, respectively,
\begin{eqnarray}
{\rm (i)} &\:& U(t_2,t_0) = \openone,\\
{\rm (ii)} &\:& U(t_2,t_0) = -
\exp \left[-i \omega_0 (t_2 - t_1)\sigma_z \right],
\nonumber 
\end{eqnarray}
which leads to the conclusion that refocusing in the physical frame
may only be achieved with identical pulses, that is, if $\varphi_j
(t)=\omega_0 t_j$.  Clearly, for the choice $\varphi_j (t)=0$, the
accumulated phase may only be disregarded in the frame rotating with
the frequency $\omega_0$.  Both choices are equally useful if
decoherence suppression becomes the primary objective in the 
open system case.


\begin{thebibliography}{68}
\bibitem[{haa()}]{haar}
\bibinfo{note}{It is worth noting that $(1/|{\cal G}|)\sum_{g \in {\cal G}}
  \sum(\cdot)$ corresponds to integration according to the invariant Haar
  measure on ${\cal G}$ in the discrete case considered here.}
\bibitem[{rem()}]{remark}
\bibinfo{note}{Note that the first pulse is assumed to occur at $t_1$, but the
  cyclicity condition with the consequent phase removal are equally satisfied
  if we start pulsing the system at the initial time $t_0$. In this case,
  however, the last pulse occurs at $t_{M-1}$. The qubit evolves freely from
  $t_{M-1}$ to $t_M$, where it should then be observed.}
\bibitem[{not()}]{note*}
\bibinfo{note}{The $\ast$-subscript is meant to stress the purely statistical
  origin of such a damping factor, somewhat reminiscent of the $T_2^\ast$
  dephasing time resulting from sample-to-sample variations in ensemble
  measurements.}
\bibitem[{sta()}]{standard}
\bibinfo{note}{The standard deviation considered here is a measure of how
  spread the results for different realizations are. It is the square root of
  the unbiased variance $s_{N-1}^2$, computed as \[ s_{N-1}^2 \equiv
  \frac{1}{N-1} \sum_{i=1}^N (x_i - \overline{x})^2\:, \] where $\overline{x}$
  is the sample mean, $x_i$ corresponds to the different sample data and $N$ is
  the total number of data. Instead, the standard deviation mentioned in
  Sect.~III is a measure of how spread the average obtained from a sample of
  the ensemble (sample mean/average value) is with respect to the average
  obtained from the full ensemble (expected value). It was used to infer the
  dependence of the sample size on the interval between pulses.}
\end{thebibliography}
\end{document}